\documentclass[aps,prl,reprint,showpacs,preprintnumbers,superscriptaddress,nobibnotes]{revtex4-2}
\pdfoutput=1
\usepackage{graphicx}
\usepackage{color}
\usepackage{orcidlink}
\usepackage{amsmath}
\usepackage{amssymb}
\usepackage{braket}
\usepackage{cleveref}
\usepackage{xspace}
\usepackage[caption=false]{subfig}
\usepackage{slashed} 
\newcommand{\cor}{\ensuremath{\mathcal{M}}\xspace}
\newcommand{\te}[1][]{\ensuremath{e^{-i H {#1}}}\xspace}
\newcommand{\tedag}[1][]{\ensuremath{e^{i H {#1}}}\xspace}
\newcommand{\z}{\ensuremath{\Delta z}\xspace}
\newcommand{\finalsite}{\ensuremath{f}\xspace}
\newcommand{\startsite}{\ensuremath{s}\xspace}
\newcommand{\W}{\ensuremath{W}\xspace}
\newcommand{\dtrotter}{\ensuremath{\tau}\xspace}
\newcommand{\Nt}{\ensuremath{N_{\dtrotter}}\xspace}
\newcommand{\dt}{\ensuremath{\Delta t}\xspace}
\newcommand{\Lcut}{\ensuremath{L_{\text{cut}}}\xspace}

\newcommand{\bjorx}{{\ensuremath{\xi}}\xspace}
\newcommand{\pdf}{\ensuremath{f_{\Psi}\left(\bjorx\right)}\xspace}
\newcommand{\pdfbare}{\ensuremath{f_{\Psi}}\xspace}
\newcommand{\antipdf}{\ensuremath{f_{\overline{\Psi}}\left(\bjorx\right)}\xspace}
\newcommand{\pdfantipdf}{\ensuremath{f_{\Psi/\overline{\Psi}}\left(\bjorx\right)}\xspace}
\newcommand{\mass}{\ensuremath{\tilde{m}}\xspace}
\newcommand{\vol}{\ensuremath{\tilde{V}}\xspace}
\newcommand{\Q}{\ensuremath{q}\xspace}
\newcommand{\Qdense}{\ensuremath{\rho}\xspace}

\newcommand{\cneg}[1]{\ensuremath{e^{-i \Tilde{\theta}_{#1}}}\xspace}
\newcommand{\cpos}[1]{\ensuremath{e^{i \Tilde{\theta}_{#1}}}\xspace}
\renewcommand{\Re}{\operatorname{Re}}
\renewcommand{\Im}{\operatorname{Im}}
\newcommand{\conj}[1]{\ensuremath{{#1}^{\dagger}}}
\newcommand{\figpdfmassdep}{fig.~4 in the main text\xspace}
\newcommand{\figpdfxdep}{fig.~3 in the main text\xspace}
\newcommand{\appendixref}[1]{Appendix~\hyperref[app:#1]{#1}~\cite{supplemental}\xspace}
\newcommand{\eqPDF}{eq.~(3) in the main text\xspace}
\newcommand{\eqH}{eq.~(2) in the main text\xspace}
\newcommand{\eqLagrangian}{eq.~(1) in the main text\xspace}
\newcommand{\eqMatrixElements}{eq.~(6) in the main text\xspace}
\usepackage[skins]{tcolorbox}
\usepackage{pgfplots} 
\pgfplotsset{compat=1.14}
\usepgfplotslibrary{fillbetween}

\usepackage{tikz}
\usetikzlibrary{external}

\usetikzlibrary{arrows,backgrounds,fit,calc,patterns,shapes}

\usetikzlibrary{plotmarks}
\pgfplotsset{compat=newest}
\usetikzlibrary{plotmarks}
\usetikzlibrary{arrows.meta}
\usepgfplotslibrary{patchplots}

\tikzset{every node/.style={sloped,allow upside down},baseline={([yshift=-0.5ex]current bounding box.center)},inner sep=.3mm,x=1cm,y=1cm}

\newcommand{\tikzlinewidth}{1.0pt}
\newcommand{\tikzmarksize}{2.0pt}

\definecolor{mycolor2}{HTML}{E40303}
\definecolor{mycolor3}{HTML}{FF8C00}
\definecolor{mycolor1}{HTML}{008026}
\definecolor{mycolor4}{HTML}{004DFF}
\definecolor{mycolor5}{HTML}{750787}
\definecolor{mycolor6}{HTML}{613915}
\definecolor{mycolor7}{HTML}{000000}
\definecolor{mycolor8}{HTML}{00C0C0}
\definecolor{mycolor9}{HTML}{000000}
\definecolor{mycolor10}{HTML}{CC8CA0}
\definecolor{mycolorblack}{rgb}{0,0,0}

\newcommand{\lightercolor}[3]{
    \colorlet{#3}{#1!#2!white}
}
\lightercolor{mycolor1}{50}{mycolor1Light}
\lightercolor{mycolor2}{50}{mycolor2Light}
\lightercolor{mycolor3}{50}{mycolor3Light}
\lightercolor{mycolor4}{50}{mycolor4Light}
\lightercolor{mycolor5}{50}{mycolor5Light}
\lightercolor{mycolor6}{50}{mycolor6Light}
\lightercolor{mycolor7}{50}{mycolor7Light}
\lightercolor{mycolor8}{50}{mycolor8Light}
\lightercolor{mycolor9}{50}{mycolor9Light}
\lightercolor{mycolor10}{50}{mycolor10Light}
\lightercolor{mycolorblack}{50}{mygray}

\definecolor{myred}{rgb}{0.478,0.016,0.012}%
\definecolor{myorange}{rgb}{0.941,0.357,0.071}%
\definecolor{myyellow}{rgb}{0.922,0.91,0.478}%
\definecolor{mygreen}{rgb}{0.400,0.612,0.086}
\definecolor{myblue}{rgb}{0.243,0.608,0.996}%
\definecolor{myviolet}{rgb}{0.353,0.137,0.549}
\definecolor{tensorblue}{rgb}{0.8,0.8,1}
\tikzset{tengreen/.style={fill=green!50!black!50}}

\tikzset{ket/.style={circle,draw=black,thick,fill=tensorblue}}
\tikzset{timeline/.style={->,line width=0.5mm,color=mygreen}}
\tikzset{spaceline/.style={->,line width=0.5mm,color=myviolet}}

\tikzset{myline1/.style={color=mycolor1, line width=\tikzlinewidth, mark size=\tikzmarksize, mark=x, mark options={solid, mycolor1}}}
\tikzset{myline2/.style={color=mycolor2, line width=\tikzlinewidth, mark size=\tikzmarksize, mark=o, mark options={solid, mycolor2}}}
\tikzset{myline3/.style={color=mycolor3, line width=\tikzlinewidth, mark size=\tikzmarksize, mark=square, mark options={solid, mycolor3}}}
\tikzset{myline4/.style={color=mycolor4, line width=\tikzlinewidth, mark size=\tikzmarksize, mark=diamond, mark options={solid, mycolor4}}}
\tikzset{myline5/.style={color=mycolor5, line width=\tikzlinewidth, mark size=\tikzmarksize, mark=triangle, mark options={solid, mycolor5}}}
\tikzset{myline6/.style={color=mycolor6, line width=\tikzlinewidth, mark size=\tikzmarksize, mark=+, mark options={solid, mycolor6}}}
\tikzset{myline7/.style={color=mycolor7, line width=\tikzlinewidth, mark size=\tikzmarksize, mark=pentagon, mark options={solid, mycolor7}}}
\tikzset{myline8/.style={color=mycolor8, dashed, line width=\tikzlinewidth, mark size=\tikzmarksize, mark=+, mark options={solid, mycolor8}}}
\tikzset{myline9/.style={mycolor9, solid, line width=\tikzlinewidth, mark size=\tikzmarksize, mark=triangle, mark options={solid, mycolor9, rotate=180}}}
\tikzset{myline10/.style={color=mycolor10, dotted, line width=\tikzlinewidth, mark size=\tikzmarksize, mark=pentagon, mark options={solid, mycolor10}}}
\tikzset{mylineblack/.style={color=mycolorblack, line width=\tikzlinewidth, mark size=\tikzmarksize, mark=triangle, mark options={solid, mycolorblack}}}

\tikzset{myphantomline1/.style={color=mycolor1Light, line width=0pt, mark size=0.0pt, mark=none}}
\tikzset{myphantomline2/.style={color=mycolor2Light, line width=0pt, mark size=0.0pt, mark=none}}
\tikzset{myphantomline3/.style={color=mycolor3Light, line width=0pt, mark size=0.0pt, mark=none}}
\tikzset{myphantomline4/.style={color=mycolor4Light, line width=0pt, mark size=0.0pt, mark=none}}
\tikzset{myphantomline5/.style={color=mycolor5Light, line width=0pt, mark size=0.0pt, mark=none}}
\tikzset{myphantomlineblack/.style={color=mycolor5Light, line width=0pt, mark size=0.0pt, mark=none}}

\tikzset{myarea1/.style={color=mycolor1Light}}
\tikzset{myarea2/.style={color=mycolor2Light}}
\tikzset{myarea3/.style={color=mycolor3Light}}
\tikzset{myarea4/.style={color=mycolor4Light}}
\tikzset{myarea5/.style={color=mycolor5Light}}
\tikzset{myarea6/.style={color=mycolor6Light}}
\tikzset{myarea7/.style={color=mycolor7Light}}
\tikzset{myarea8/.style={color=mycolor8Light}}
\tikzset{myarea9/.style={color=mycolor9Light}}
\tikzset{myarea10/.style={color=mycolor10Light}}
\tikzset{myarea11/.style={color=mycolor11Light}}
\tikzset{myarea12/.style={color=mycolor12Light}}
\tikzset{myareablack/.style={color=mygray}}

\pgfplotsset{
  /pgfplots/error bar legend/.style={
    legend image code/.code={
        \draw[sharp plot,mark=-,mark repeat=2,mark phase=1,##1]
        plot coordinates { (0.3cm, -0.15cm) (0.3cm,0cm) (0.3cm, 0.15cm) };%
        \draw[mark repeat=2,mark phase=2,##1]
        plot coordinates {(0cm,0cm) (0.3cm,0cm) (0.6cm,0cm)};%
        }}}

\pgfplotsset{
  /pgfplots/error band legend/.style={
    legend image code/.code={
        \draw[fill=#1, draw=none] (-0cm,-0.15cm) rectangle ++(0.6cm,0.3cm);
        \draw[sharp plot,mark=-##1] (0cm,0cm) -- (0.6cm,0cm);%
        }}}
\begin{document}

\title{
Parton Distribution Functions in the Schwinger model from Tensor Network States
}
\author{Mari~Carmen~Ba\~{n}uls\orcidlink{0000-0001-6419-6610}}
\affiliation{Max Planck Institute of Quantum Optics, Garching, Germany}

\author{Krzysztof~Cichy\orcidlink{0000-0002-5705-3256}}
\affiliation{Faculty of Physics and Astronomy, Adam Mickiewicz University, Pozna\'{n}, Poland}

\author{C.-J.~David~Lin\orcidlink{0000-0003-3743-0840}}
\affiliation{Institute of Physics, National Yang Ming Chiao Tung University, Hsinchu, Taiwan}
\affiliation{Center for Theoretical and Computational Physics, National Yang Ming Chiao Tung University, Hsinchu, Taiwan}
\affiliation{Centre for High Energy Physics, Chung-Yuan Christian University, Taoyuan, Taiwan}

\author{Manuel~Schneider\orcidlink{0000-0001-9348-8700}}
\email{manuel.schneider@nycu.edu.tw}
\affiliation{Institute of Physics, National Yang Ming Chiao Tung University, Hsinchu, Taiwan}
\affiliation{Center for Theoretical and Computational Physics, National Yang Ming Chiao Tung University, Hsinchu, Taiwan}

\date{April 19, 2025}

\begin{abstract}
Parton distribution functions (PDFs) describe the inner, non-perturbative structure of hadrons. Their computation involves matrix elements with a Wilson line along a direction on the light cone, posing significant challenges in Euclidean lattice calculations, where the time direction is not directly accessible. We propose implementing the light-front Wilson line within the Hamiltonian formalism using tensor network techniques. The approach is demonstrated in the massive Schwinger model (quantum electrodynamics in 1+1 dimensions), a toy model that shares key features with quantum chromodynamics. We present accurate continuum results for the fermion PDF of the vector meson at varying fermion masses, obtained from first-principle calculations directly in Minkowski space. Our strategy also provides a useful path for quantum simulations and quantum computing.
\end{abstract}

\preprint{\doi{10.48550/arXiv.2504.07508}}

\maketitle

\phantomsection
\addcontentsline{toc}{section}{Introduction}
{\sl Introduction ---}
Investigating the internal structure of protons and neutrons (nucleons), the nuclei's building blocks, is of paramount importance for modern physics. Significant progress is expected in the next years, with increased experimental efforts, such as the Electron-Ion Collider under construction at the Brookhaven National Laboratory~\cite{NAP25171}. However, a full understanding of experimental developments requires the support of a robust theoretical description, embodied in different kinds of partonic functions. These functions describe the momentum- and coordinate-space distributions of partons, i.e. quarks and gluons, inside the nucleon.
The simplest, yet essential, partonic functions are parton distribution functions (PDFs). Even though, in principle, they can be obtained directly from quantum chromodynamics (QCD), theoretical access to them is limited by their non-perturbative nature.

In the parton picture, the PDF describes the probability of finding a constituent in a hadron with a fraction $\xi$ of the momentum. Computing the PDF on the lattice involves the calculation of matrix elements with operators separated along a light-front direction. However, for practical reasons, numerical calculations of lattice QCD have been predominantly performed using the path integral formalism, which requires the formulation of the theory in Euclidean spacetime. Thus, it gives no {\it direct} access to light-front dynamics. In the past two decades, novel approaches have been proposed~\cite{Liu:1999ak, Detmold:2005gg, Braun:2007wv,Davoudi:2012ya,Ji:2013dva,Ma:2014jla,Radyushkin:2017cyf, Chambers:2017dov, Detmold:2021uru,Shindler:2023xpd} which calculate intermediate Euclidean objects that can be matched to their Minkowski counterparts. Although tremendous theoretical and numerical progress has been made, much work is still needed to achieve an accurate determination of PDFs from Euclidean lattice QCD~\cite{Cichy:2018mum,Radyushkin:2019mye,Ji:2020ect,Constantinou:2020pek,Cichy:2021lih,Cichy:2021ewm,Gao:202489}. Therefore, it remains desirable to have first-principle theoretical determinations of PDFs with computations performed directly in Minkowski space. Such calculations can, in principle, be achieved in the Hamiltonian formalism. However, previous works that extracted PDFs in the Hamiltonian approach~\cite{MoPerry,BERGKNOFF1977215,KROGER199858} were based on few-fermion approximations or small momentum lattices and relied on exact diagonalization. These techniques are not scalable and do not allow one to control the systematic errors reliably.

Tensor network (TN) methods~\cite{Verstraete2008,SCHOLLWOCK201196,Orus2014annphys,Bridgeman_2017,silvi2019tensor,Cirac2021,Okunishi2022,Banuls_TNA_RouteMap} provide an efficient entanglement-based ansatz for quantum many-body problems that could potentially overcome these limitations. They have been very successful in describing equilibrium states of low-dimensional lattice gauge theories (LGTs)~\cite{banuls2020ropp,banuls2020epjd,meurice2022trg,Kadoh:2022Ia,felser2020u1,robaina2021z3,magnifico2021qed3d,emonts2023z2,kelman2024gpeps,Akiyama_2024} and even achieved the most precise continuum extrapolations for some cases. Tensor network algorithms in the Hamiltonian formalism can simulate real-time evolution of states with moderate entanglement~\cite{Verstraete2008,SCHOLLWOCK201196,Paeckel2019}. Nevertheless, studies of LGTs in this regard have been less common (see references in~\cite{banuls2020ropp,banuls2020epjd}), and have not yet accomplished a systematic extrapolation to the continuum limit.

Here, we close this gap and
demonstrate that TN methods are also suitable to determine real-time properties of gauge theories in the continuum. Our study shows that PDFs can be calculated with controlled systematic errors by a time evolution of tensor network states, in contrast to previous approaches in the Hamiltonian framework~\cite{MoPerry,BERGKNOFF1977215,KROGER199858}, in which uncontrolled systematic effects are present. In particular, the entanglement entropy grows during the time evolution of a state. While this can, in principle, challenge the applicability of tensor network methods, we observe that it is not a practical limitation for the calculation of PDFs in the Schwinger model.

Complementary to TN methods, quantum computation and quantum simulation are being explored in recent years as technologies that could potentially overcome the exponential complexity barrier of quantum many-body problems. Their applicability to LGTs has motivated multiple studies~\cite{banuls2020epjd,davoudi2020,Funcke2023,PRXQuantum.4.027001,halimeh2023coldatom,dimeglio2023quantum}, including the possibility of extracting PDFs~\cite{PRXQuantum.4.027001,Pedernales2014,Lamm2019,Lamm2020,Echevarria2021,Pisarski2022,Kreshchuk2021,Kreshchuk2022,Tianyin2022,Wenyang2022,Grieninger2024}. Our approach in the Hamiltonian formalism could also be implemented on quantum hardware.

In this work, we propose a strategy for accessing light-cone dynamics directly in Minkowski space~\footnote{The method has been presented at the 41st International Symposium on Lattice Field Theory~\cite{Schneider:2025gL}. Similar ideas have since been applied to a purely fermionic model~\cite{2025arXiv250109738K}.}\nocite{Schneider:2025gL,2025arXiv250109738K}. Its feasibility is demonstrated by calculating the PDF defined in \cref{eq:PDF} of the vector meson in the massive Schwinger model for different values of the fermion mass with TN methods. We show that errors can be systematically controlled, allowing for a precise extrapolation to the continuum limit.

\phantomsection
\addcontentsline{toc}{section}{Model}
{\sl Model ---}
The Schwinger model, or quantum electrodynamics (QED) in 1+1 dimensions~\cite{SchwingerOriginal,SchwingerModelMassive}, shares key features with QCD, including dynamical mass generation, confinement, and asymptotic freedom. Consequently, it has been widely used to study QCD-like phenomena and to test new methods in high-energy physics. The QED Lagrange density is given by~\footnote{\label{fn:conventions}We use Einstein summation convention, the slashed notation $\slashed{A} = \gamma^\mu A_\mu$ with $\gamma^0 = \sigma_z = \begin{pmatrix}1&0\\0&-1\end{pmatrix}$ and $\gamma^1 = i \sigma_y = \begin{pmatrix}0&1\\-1&0\end{pmatrix}$, $\overline{\Psi} = \Psi^\dagger \gamma^0$, and $\sigma^{\pm} =\frac{1}{2}\left(\sigma^{x}\pm i\sigma^{y}\right) = \frac{1}{2} \begin{pmatrix}    0	&	1 \pm 1\\  1 \mp 1	&	0  \end{pmatrix}$.}
\begin{equation}
	\mathcal{L} = \overline{\Psi} (i \slashed{\partial}- g \slashed{A }-m )\Psi-\frac{1}{4}F_{\mu\nu}F^{\mu \nu}.
	\label{eq:Lagrangian}
\end{equation}
Here, $A_\mu$ are $U(1)$ gauge fields with field strength tensor $F_{\mu\nu} = \partial_\mu A_\nu - \partial_\nu A_\mu$, $m$ is the fermion mass, and $g$ the coupling strength. The Hamiltonian formulation follows from a Legendre transformation. We discretize space on a finite lattice of $N$ sites, and employ staggered fermions~\cite{staggeredFermions} in the temporal gauge $A^0 = 0$. Using a Jordan-Wigner transformation, which maps the fermionic degrees of freedom to spins, and integrating out the gauge variables with the help of Gauss's law, the following Hamiltonian is derived (see~\cite{Schwinger_massSpectrum,Hamer1997,Banks1976,Schwinger_QC} and \appendixref{A}):\nocite{Crewther1980,henneaux1992quantization,Zohar2015}
\begin{align}
	H =& x\sum_{n=0}^{N-2} \left [ \sigma_n^+\sigma_{n+1}^- +  \sigma_n^-\sigma_{n+1}^+ \right ]+\frac{\mu}{2}\sum_{n=0}^{N-1} \left [ 1 + (-1)^n \sigma_n^z \right ] \nonumber\\
	&+\sum_{n=0}^{N-2} \left [\sum_{k=0}^n \left(\frac{1}{2}\left((-1)^k+\sigma_k^{z}\right)  + \Q_k \right) \right ]^2.
	\label{eq:H}
\end{align}
It is equivalent to the Schwinger model in the thermodynamic and continuum limit, with the latter approached as the lattice spacing $a$ goes to zero, or equivalently, when $x \equiv \frac{1}{a^2g^2} \rightarrow \infty$. The dimensionless fermion mass $\mu = \frac{2m}{ag^2} - \frac{1}{4}$ includes an additive renormalization to enhance convergence to the continuum limit~\cite{Schwinger_latticeMass,MassRenormalizationNumerical,MassRenormalizationNumerical_PoS}. Each constant $\Q_k$ represents the presence of a static charge at site $k$.

\phantomsection
\addcontentsline{toc}{section}{Implementation}
{\sl Implementation ---}
For a hadron with momentum $P$, the unpolarized PDF $f_{\Psi}(\xi)$ for the fermionic parton described by the spinor field $\Psi$ is defined as~\cite{collins_2011}
\begin{equation}
	\pdf = \int \frac{dz^{-}}{4\pi} e^{-i \bjorx P^+ z^-}\hspace{-1.61365pt} \Braket{P|\overline{\Psi} (z^-) \W_{(z^-,0)} \gamma^+ \Psi(0)|P} .
	\label{eq:PDF}
\end{equation}
We denote the light-front components of a Lorentz vector $v = (v^{0}, v^{1}, \ldots, v^{d})$ in $d+1$ dimensions as $v^{\pm} = (v^{0}\pm v^{d})/\sqrt{2}$.  Here, $\xi$ is the Bjorken scaling variable, which, in the parton model, is equivalent to the fraction of the hadron momentum carried by the fermion, and $\W_{(z_{2},z_{1})}$ is a Wilson line connecting spacetime points $z_{1}$ and $z_{2}$ along the light-front direction.

We use the matrix product state (MPS) TN ansatz~\cite{Verstraete2008,SCHOLLWOCK201196} to approximate the meson eigenstate~\cite{Schwinger_massSpectrum}, as well as the time-dependent vector resulting from the time evolution needed for the implementation of the Wilson line, in the physical subspace in which the Hamiltonian of \cref{eq:H} is expressed.

\begin{figure}[tb]
	\centering
	\subfloat[\label{sfig:zigzag}]{%
		\begin{minipage}{.45\columnwidth}
			\def\figWidth{.7\textwidth}
\begin{tikzpicture}
	\begin{axis}[
		width=\figWidth,
		height=\figWidth,
		scale only axis,
		scale only axis,
		xmin=0,
		xmax=2*3+2,
		xlabel style={font=\color{white!15!black},yshift=-1ex},
		xlabel={position [lattice units]},
		ymin=-1,
		ymax=2*3+2-1,
		ylabel style={font=\color{white!15!black},xshift=1ex},
		ylabel={time [lattice units]},
		axis background/.style={fill=white},
		xmajorgrids,
		ymajorgrids,
		legend style={at={(1.03,1)}, anchor=south west, legend cell align=left, align=left, draw=white!15!black},
		ylabel style={{rotate=-90}},
		scaled ticks=false,
		tick label style={/pgf/number format/fixed},
		xtick={0,2,...,10},
		ytick={0,2,...,10},
        minor tick num=1,
        xminorgrids,
		yminorgrids,
		xticklabel style={yshift=-1ex},
		yticklabel style={xshift=-1ex}
		]
		\node[ket,tengreen] (site0) at (1,0) {\tiny $\sigma^-$};
		\node[] (time0) at (0*2+1,1*2) {};
		\node[] (site1) at (1*2+1,1*2) {};
		\node[] (time1) at (1*2+1,2*2) {};
		\node[] (site2) at (2*2+1,2*2) {};
		\node[] (time2) at (2*2+1,3*2) {};
		\node[ket,tengreen] (site3) at (3*2+1,3*2) {\tiny $\sigma^+$};
		\draw[black,dashed] (site0) -- (site3) node[midway,below=1pt,opacity=1,fill=white] {light cone};
        \draw[timeline] (site0) -- (time0);
		\draw[spaceline] (time0) -- (site1);
        \draw[timeline] (site1) -- (time1);
		\draw[spaceline] (time1) -- (site2);
        \draw[timeline] (site2) -- (time2);
		\draw[spaceline] (time2) -- (site3);
	\end{axis}
\end{tikzpicture}%
		\end{minipage}
	}\hfill
	\subfloat[\label{sfig:light-cone}]{%
		\begin{minipage}{.52\columnwidth}
			\includegraphics[width=\textwidth]{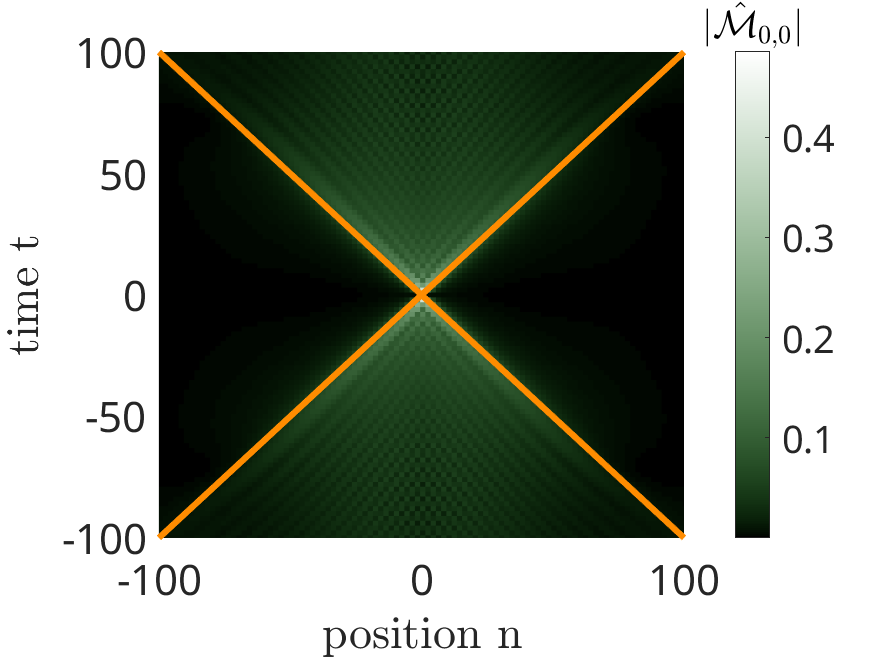}
			\vfill
		\end{minipage}
	}
	\caption{Light-cone matrix elements. a) Operators $\sigma^-$ and $\sigma^+$ from \cref{eq:MatrixElements} separated along a light front are calculated on the lattice by subsequent steps of a spatial evolution of the Wilson line (violet horizontal arrows) and time evolution steps (green vertical arrows). b) The light-cone structure that emerges in matrix elements; shown is the modulus of an even-to-even matrix element $\hat{\mathcal{M}}_{0,0} =$ $\Bra{h} \tedag[\frac{\dt}{2} t] \prod_{k<\z}\left(i\sigma_{k}^{z}\right) \sigma_{\z}^{+} \te[_{0} \frac{\dt}{2} t] \prod_{k'<0}\left(-i\sigma_{k'}^{z}\right) \sigma_{0}^{-} \Ket{h}$ for a light mass $\frac{m}{g} = 0.125$. $\hat{\mathcal{M}}_{0,0}$ is similar to $\mathcal{M}_{0,0}$ in \cref{eq:MatrixElements}, but with a fixed static charge to illustrate the emerging light-cone structure without enforcing a light-front direction by moving the static charge. Orange lines indicate the light-front $+$ and $-$ directions.}
	\label{fig:zigzag}
\end{figure}
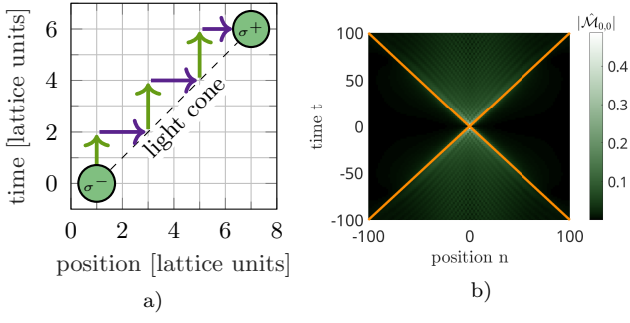

The matrix element in \cref{eq:PDF} 
involves a Wilson line along a direction on the light cone, connecting two fermionic operators. The latter, in the spin formulation discussed above, correspond to $\sigma^{\pm}$ operators preceded by a Jordan-Wigner string of $\sigma^z$. On the lattice, we approximate the light front by a stepwise evolution, as illustrated in \cref{sfig:zigzag}. Each evolution by a timestep $\dt$ is followed by a spatial evolution of the Wilson line by two lattice sites. Setting $\dt = \frac{1}{x}$ corresponds to a speed of light $c=1$, ensuring that the light cone is approached in the continuum limit~\footnote{The rescaling factor $\frac{a g^2}{2}$ from \appendixref{A} is included here.}. The light-cone structure that emerges is illustrated in \cref{sfig:light-cone}. In the temporal gauge $A^0 = 0$, the Wilson line in time direction becomes the identity operator. Consequently, the vertical lines in \cref{sfig:zigzag} correspond solely to the application of a time evolution operator $e^{-iH \dt}$.

The steps along the space direction, 
instead, act non-trivially on the gauge (link) variables, with the effect of shifting the electric flux term of the Hamiltonian by one unit on one link at a time~\cite{Banks1976,Irving1983}. This is exactly like having a pair of charges around that link. Thus, the string can be absorbed in the time evolution by introducing a static charge $\Q_k$ that \textit{ends} the Wilson line and moves along the light cone (see \appendixref{B}). Horizontal arrows in \cref{sfig:zigzag} starting at a point $a$ with endpoint $b$ thus correspond to decreasing the static charge $\Q_a$ and increasing $\Q_b$ by one unit each~\footnote{Notice that, at the initial time, a single (positive) static charge sits on the same vertex as the initial fermionic operator, thus canceling the change in electric flux introduced by the latter.}. This formulation is most suitable for working in the physical subspace, where gauge degrees of freedom have been integrated out. Similar ideas that include static charges have been used to calculate Wilson loops in the Hamiltonian formalism~\cite{Echevarria2021,Pisarski2022}.

With these techniques, the PDF in \cref{eq:PDF} can be evaluated on the lattice and becomes for a hadron of mass $M$ at rest:
\begin{equation}
	\pdf = \frac{NM}{8\pi x} \sum_{\z=0,2,4,...,N} e^{-i \bjorx \frac{M \z}{2x}} \mathcal{M}(\z)
	\label{eq:PDF_discrete}.
\end{equation}
If the system is translationally invariant, the matrix elements $\mathcal{M}$ will depend only on the distance between the end points of the Wilson line. Due to the staggered fermions, the Dirac structure leads, however, to four contributions to the matrix elements $\mathcal{M}(\z)$, corresponding to all combinations between even (subscript $f$ or $s=0$) and odd (subscript $f$ or $s=1$) start and end sites~\footnote{From here on, we use a notation where site 0 is an even site in the middle of the lattice.}:
\begin{align}
	\mathcal{M}&(\z) = \mathcal{M}_{0,0} - \mathcal{M}_{0,1} - \mathcal{M}_{1,0} + \mathcal{M}_{1,1} ,\label{eq:MatrixElement}\\
	\mathcal{M}&_{\finalsite,\startsite}(\z) = \Bra{h} \tedag[\frac{\dt}{2} \z] \prod_{k<\z+\finalsite}\left(i\sigma_{k}^{z}\right) \sigma_{\z+\finalsite}^{+} \label{eq:MatrixElements}\\
	&\te[_{\z-1} \dt] \dots \te[_{3} \dt]  \te[_{1} \dt] \prod_{k'<\startsite}\left(-i\sigma_{k'}^{z}\right) \sigma_{\startsite}^{-} \Ket{h}_c .\nonumber
\end{align}
The subscript $c$ denotes the subtraction of vacuum contributions from matrix elements~\cite{Schneider:2025gL,collins_2011}: $\braket{h|\hat{O}|h}_c = \braket{h|\hat{O}|h} - \braket{0|\hat{O}|0}$. The ket $\Ket{h}$ corresponds to the first excited state of $H$, a vector meson at rest, with mass $M = \Braket{h|H|h}_c$~\cite{Schwinger_massSpectrum}. Hence, the first evolution operator simply introduces a phase $e^{iM \frac{\dt}{2}}$. Notice that for the intermediate time-evolution operators $e^{-iH_k \dt}$, the subindex $k$ in the Hamiltonian indicates the inclusion of a single positive static charge at position $k$. For $\z<0$, the evolution in the operator is implemented in a similar way. In this case, we start with a spatial evolution in the negative direction, followed by a time evolution with a negative time step $-\dt$. \Cref{sfig:light-cone} 
illustrates the light-cone and the spread of correlations at different times and distances, in the absence of the light-front Wilson line, and confirms that the speed of light is $c=1$.

We use the variational optimization described in~\cite{Schwinger_massSpectrum} to obtain an MPS approximation to the ground state $\ket{0}$ and the first excited state in the chargeless sector $\ket{h}$ of \cref{eq:H} for a finite system with open boundary conditions. For the time evolution, we use a second-order Suzuki-Trotter decomposition~\cite{Trotter,Suzuki1990FractalDO}, and divide each time step into $\Nt = \frac{\dt}{\dtrotter}$ smaller evolution steps that can be applied efficiently to the MPS via the tMPS method~\cite{Verstraete2008,SCHOLLWOCK201196,Paeckel2019}. We closely follow the approach in~\cite{Schwinger_thermalEvolution}, where the time evolution operator for the electric component of the Hamiltonian, expressed as a matrix product operator (MPO), is truncated to allow for only a change in the electric flux by at most $L_\text{cut} = 10$. By varying $L_\text{cut}$, we confirm that the results do not depend on the cutoff (see \appendixref{D}).

We set the bond dimension of the initial MPS (i.e., the one used to approximate the eigenstate) and that of the MPS during the time evolution equally to a value $D$. For all the analyzed cases, we checked that our results show no significant dependence for $D \ge 40$, and thus we choose $D = 80$. The convergence with such moderate values of the bond dimensions indicates that the ansatz is remarkably adequate to capture the properties of this problem. We observe that a large Trotter step introduces a phase error in the matrix elements, causing a shift of the fermion PDF towards smaller \bjorx. With the choice $\Nt = 100$, this effect remains insignificant compared to other errors in our simulations. We observe that the dominant error contributions in regions where the PDF is sizable (e.g.\ $\bjorx \in(0.3,0.7)$ at $\mass=10$) arise from the finite lattice spacing and the finite-volume effects, the latter dominating close to $\bjorx=0.5$. A detailed error analysis is provided in \appendixref{D}.

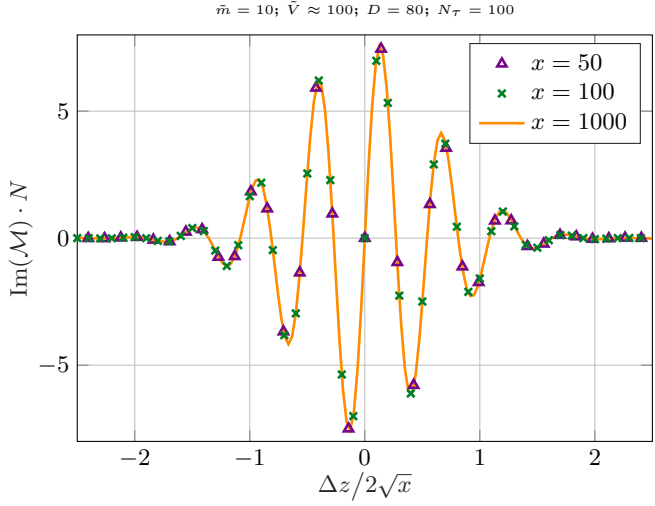
\begin{figure}[tbp]
	\centering
	\def\figWidth{.88*\columnwidth}
	\def\figHeight{.3*\textwidth}
%
%
\begin{tikzpicture}

\begin{axis}[%
width=\figWidth,
height=\figHeight,
at={(0\figWidth,0\figHeight)},
scale only axis,
xmin=-2.5,
xmax=2.5,
xlabel style={font=\color{white!15!black}},
xlabel={$\z \big/ 2\sqrt{x}$},
ymin=-8,
ymax=8,
ylabel style={font=\color{white!15!black}},
ylabel={$\Im(\cor) \cdot N$},
axis background/.style={fill=white},
title style={font=\bfseries},
title={\tiny $\mass = 10$; $\vol \approx 100$; $D=80$; $\Nt = 100$},
xmajorgrids,
ymajorgrids,
legend style={at={(0.98,0.98)}, anchor=north east, legend cell align=left, align=left, draw=white!15!black}, 
ylabel style={{rotate=-90}}, scaled ticks=false, tick label style={/pgf/number format/fixed},
xticklabel style={yshift=-.5ex},
yticklabel style={xshift=-.5ex}
]
\addplot [myline5, only marks]
  table[row sep=crcr]{%
-2.54558441227157	0.00232349876776941\\
-2.40416305603426	-0.00345354110154798\\
-2.26274169979695	-0.0101976546600339\\
-2.12132034355964	0.0160797186650944\\
-1.97989898732233	0.0401164936734924\\
-1.83847763108502	-0.0682508951108527\\
-1.69705627484771	-0.12994878330765\\
-1.5556349186104	0.246592079163153\\
-1.41421356237309	0.341851675879871\\
-1.27279220613579	-0.741860880089953\\
-1.13137084989848	-0.716902551227659\\
-0.989949493661166	1.83400086829115\\
-0.848528137423857	1.1606823601539\\
-0.707106781186547	-3.68041548219369\\
-0.565685424949238	-1.35963321982575\\
-0.424264068711929	5.91530007528865\\
-0.282842712474619	0.961744664135685\\
-0.14142135623731	-7.50057332600353\\
0	-1.13619982776725e-13\\
0.14142135623731	7.44724807321515\\
0.282842712474619	-0.951414654359895\\
0.424264068711929	-5.78733699474754\\
0.565685424949238	1.3313972483098\\
0.707106781186547	3.54623799973663\\
0.848528137423857	-1.12332787850394\\
0.989949493661166	-1.73891142086817\\
1.13137084989848	0.684078803205602\\
1.27279220613579	0.691018272134207\\
1.41421356237309	-0.320845008345253\\
1.5556349186104	-0.225229221993908\\
1.69705627484771	0.119601667378312\\
1.83847763108502	0.0611029894386152\\
1.97989898732233	-0.0360534748601554\\
2.12132034355964	-0.0142243976618189\\
2.26274169979695	0.00898768528258019\\
2.40416305603426	0.00340268506527603\\
};
\addlegendentry{$x=50$}

\addplot [myline1, only marks]
  table[row sep=crcr]{%
-2.5	0.00159851924837943\\
-2.4	-0.00358030790988884\\
-2.3	-0.00968975355920187\\
-2.2	-0.00390952129160433\\
-2.1	0.0219438498094239\\
-2	0.0408347656203673\\
-1.9	-0.00241731411752107\\
-1.8	-0.107174392111173\\
-1.7	-0.13041846176761\\
-1.6	0.0880820593422123\\
-1.5	0.397603356909738\\
-1.4	0.279913769386882\\
-1.3	-0.49063515924158\\
-1.2	-1.09460882336512\\
-1.1	-0.277907570990493\\
-1	1.65252271830007\\
-0.9	2.18167296249031\\
-0.8	-0.471405203896476\\
-0.7	-3.81877354927859\\
-0.6	-2.96141935181246\\
-0.5	2.546670899693\\
-0.4	6.20424054188569\\
-0.3	2.28467017140179\\
-0.2	-5.36589456109471\\
-0.1	-7.00646530672796\\
0	-2.65623688225602e-14\\
0.1	6.98053605388359\\
0.2	5.32498936912277\\
0.3	-2.2618251214808\\
0.4	-6.11191228179803\\
0.5	-2.4931272337924\\
0.6	2.9019932763187\\
0.7	3.71763949751341\\
0.8	0.448754463946059\\
0.9	-2.11353130400834\\
1	-1.58721198575306\\
1.1	0.272587511236122\\
1.2	1.04646527548781\\
1.3	0.463239528627872\\
1.4	-0.267752853625209\\
1.5	-0.374484469643035\\
1.6	-0.0810526592685681\\
1.7	0.122545169049254\\
1.8	0.0993106599884909\\
1.9	0.00201914245895223\\
2	-0.037753432459906\\
2.1	-0.0200038602628575\\
2.2	0.00350898188992918\\
2.3	0.00869864634037236\\
2.4	0.00348253528608653\\
};
\addlegendentry{$x=100$}

\addplot [myline3, mark=none]
  table[row sep=crcr]{%
-2.49819935153302	0.00550475556911521\\
-2.46657657493134	0.00579508768016205\\
-2.43495379832965	0.00350622615483725\\
-2.40333102172797	0.0003340384209163\\
-2.37170824512628	-0.00320952211367056\\
-2.3400854685246	-0.00528170547402899\\
-2.30846269192292	-0.00672940661336275\\
-2.27683991532123	-0.00658871209648591\\
-2.24521713871955	-0.00528418244110801\\
-2.21359436211787	-0.00271363930462278\\
-2.18197158551618	0.00204932910278965\\
-2.1503488089145	0.0100848939939799\\
-2.11872603231281	0.0190443191546093\\
-2.08710325571113	0.0279338033926352\\
-2.05548047910945	0.0364963228848258\\
-2.02385770250776	0.0411674198145064\\
-1.99223492590608	0.0415941432387374\\
-1.9606121493044	0.034359959853302\\
-1.92898937270271	0.0198345415642905\\
-1.89736659610103	-0.00286021207637521\\
-1.86574381949934	-0.0337057825102992\\
-1.83412104289766	-0.0678030458021819\\
-1.80249826629598	-0.10264548106892\\
-1.77087548969429	-0.130313384224175\\
-1.73925271309261	-0.143208929432329\\
-1.70762993649092	-0.137132861384746\\
-1.67600715988924	-0.103850231734007\\
-1.64438438328756	-0.0448068843482199\\
-1.61276160668587	0.0395065293611829\\
-1.58113883008419	0.141189074447764\\
-1.54951605348251	0.247760961286271\\
-1.51789327688082	0.341433017537394\\
-1.48627050027914	0.403917175604837\\
-1.45464772367745	0.41427016038399\\
-1.42302494707577	0.357931569228119\\
-1.39140217047409	0.226179343931217\\
-1.3597793938724	0.0235042826750477\\
-1.32815661727072	-0.235365209288772\\
-1.29653384066904	-0.51981551584689\\
-1.26491106406735	-0.790061909574758\\
-1.23328828746567	-0.996587647393606\\
-1.20166551086398	-1.09181371974412\\
-1.1700427342623	-1.03543459543926\\
-1.13841995766062	-0.804016080859198\\
-1.10679718105893	-0.399141945979226\\
-1.07517440445725	0.149695666881335\\
-1.04355162785557	0.782647312923548\\
-1.01192885125388	1.4133413417128\\
-0.980306074652198	1.9415191589411\\
-0.948683298050514	2.2672550818906\\
-0.91706052144883	2.30030498596262\\
-0.885437744847146	1.98834827867949\\
-0.853814968245462	1.32409331888705\\
-0.822192191643779	0.355467123616225\\
-0.790569415042095	-0.811507841293984\\
-0.758946638440411	-2.02849503248638\\
-0.727323861838727	-3.10949527971556\\
-0.695701085237043	-3.87649259458869\\
-0.66407830863536	-4.16708451812513\\
-0.632455532033676	-3.88108988719005\\
-0.600832755431992	-2.99399039612285\\
-0.569209978830308	-1.57256779475093\\
-0.537587202228625	0.224872012556746\\
-0.505964425626941	2.16761078654273\\
-0.474341649025257	3.98013756931701\\
-0.442718872423573	5.37817178016617\\
-0.411096095821889	6.12446355214138\\
-0.379473319220206	6.05470853980658\\
-0.347850542618522	5.12446973113318\\
-0.316227766016838	3.41280813825928\\
-0.284604989415154	1.12346065888087\\
-0.25298221281347	-1.43979879614161\\
-0.221359436211787	-3.9216414324122\\
-0.189736659610103	-5.95312226137566\\
-0.158113883008419	-7.23271028782641\\
-0.126491106406735	-7.5511422810705\\
-0.0948683298050514	-6.8357225108202\\
-0.0632455532033676	-5.17323291700953\\
-0.0316227766016838	-2.782091781963\\
0	-5.08592874766671e-13\\
0.0316227766016838	2.78124415021305\\
0.0632455532033676	5.16951206978176\\
0.0948683298050514	6.82904923867923\\
0.126491106406735	7.54308938123316\\
0.158113883008419	7.21948717992491\\
0.189736659610103	5.94146348027722\\
0.221359436211787	3.90912614348294\\
0.25298221281347	1.43458004036983\\
0.284604989415154	-1.12270797072227\\
0.316227766016838	-3.40234905036042\\
0.347850542618522	-5.10702534627004\\
0.379473319220206	-6.03328220650113\\
0.411096095821889	-6.10114452412289\\
0.442718872423573	-5.35743514122653\\
0.474341649025257	-3.96254923471379\\
0.505964425626941	-2.15787741277711\\
0.537587202228625	-0.224139286133722\\
0.569209978830308	1.56491814484629\\
0.600832755431992	2.97771166098835\\
0.632455532033676	3.85852205580861\\
0.66407830863536	4.14116542961862\\
0.695701085237043	3.8496643040476\\
0.727323861838727	3.08780287528056\\
0.758946638440411	2.01253877684913\\
0.790569415042095	0.805291287268274\\
0.822192191643779	-0.353394028556098\\
0.853814968245462	-1.31303720193272\\
0.885437744847146	-1.9698295617824\\
0.91706052144883	-2.27755064883732\\
0.948683298050514	-2.24313168674243\\
0.980306074652198	-1.91971605689567\\
1.01192885125388	-1.39442113311037\\
1.04355162785557	-0.769412537575365\\
1.07517440445725	-0.143117551694656\\
1.10679718105893	0.400405036570676\\
1.13841995766062	0.801223199995333\\
1.1700427342623	1.02930201717637\\
1.20166551086398	1.08489950041368\\
1.23328828746567	0.989835681028164\\
1.26491106406735	0.78504410235528\\
1.29653384066904	0.518547467737635\\
1.32815661727072	0.236831143149939\\
1.3597793938724	-0.0192864240867692\\
1.39140217047409	-0.220108788158215\\
1.42302494707577	-0.34988867700159\\
1.45464772367745	-0.405964232311756\\
1.48627050027914	-0.395335587593259\\
1.51789327688082	-0.33425562324452\\
1.54951605348251	-0.241598279007956\\
1.58113883008419	-0.136326217216463\\
1.61276160668587	-0.0358266000273438\\
1.64438438328756	0.0465812557611716\\
1.67600715988924	0.105607532035219\\
1.70762993649092	0.136190377676645\\
1.73925271309261	0.142504153301507\\
1.77087548969429	0.128420300640075\\
1.80249826629598	0.101155571068166\\
1.83412104289766	0.0672480726189312\\
1.86574381949934	0.0327550509333\\
1.89736659610103	0.00172463954074126\\
1.92898937270271	-0.0218110885918342\\
1.9606121493044	-0.0367627759447157\\
1.99223492590608	-0.0428901097644327\\
2.02385770250776	-0.0429481259840186\\
2.05548047910945	-0.0375602217008961\\
2.08710325571113	-0.0294284661470817\\
2.11872603231281	-0.02026285611321\\
2.1503488089145	-0.0119672744294837\\
2.18197158551618	-0.00442447563331626\\
2.21359436211787	0.000272371989083531\\
2.24521713871955	0.00306704694547429\\
2.27683991532123	0.00390237187467775\\
2.30846269192292	0.00278355783129142\\
2.3400854685246	0.00184727897803081\\
2.37170824512628	-8.20025212985853e-06\\
2.40333102172797	-0.000857507859774839\\
2.43495379832965	-0.00119934066828852\\
2.46657657493134	-0.0067404663286153\\
2.49819935153302	-0.0217026448859263\\
};
\addlegendentry{$x=1000$}

\end{axis}

\end{tikzpicture}%
	\caption{Imaginary part of matrix element. The values for $x = 1000$ are discrete but shown as a line for clarity.}
	\label{fig:matrixElement}
\end{figure}

\begin{figure}[tbp]
	\centering
	\def\figWidth{.88*\columnwidth}
	\def\figHeight{.3*\textwidth}
	\input{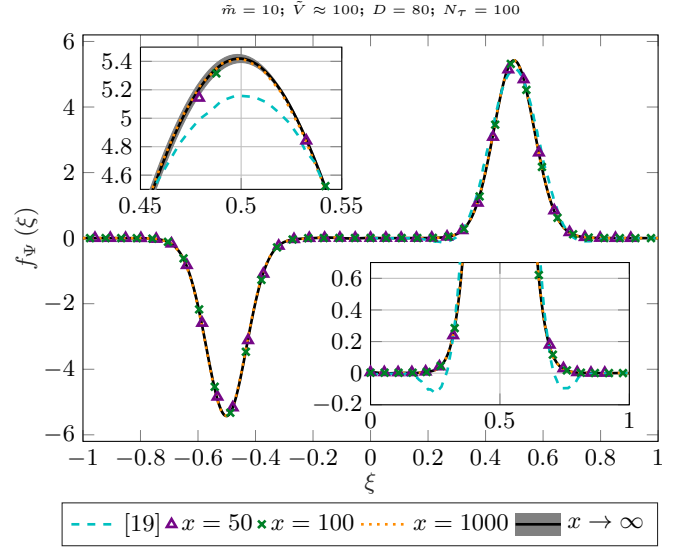}
	\caption{PDF for different lattice spacings, obtained by a discrete Fourier transform. The values for $x=1000$ are shown as a dotted line for clarity. Solid line with error bands: extrapolation to the continuum limit at fixed volume. Dashed line: results from~\cite{MoPerry} (normalized).}
	\label{fig:PDF_x_dependence}
\end{figure}

\phantomsection
\addcontentsline{toc}{section}{Results}
{\sl Results ---}
The real part of the matrix element $\mathcal{M}(\z)$ in \cref{eq:MatrixElements} can be shown to be symmetric, the imaginary part antisymmetric with respect to $\bjorx = 0$, see \appendixref{C}. We observe that the real part is several orders of magnitude smaller than the imaginary part, making its contribution to the PDF negligible. As a result, $\pdf$ is antisymmetric. The imaginary part exhibits oscillatory behavior and vanishes for large $|\z|$. \Cref{fig:matrixElement} shows the imaginary part of the matrix element for a fixed fermion mass $\mass \equiv \frac{m/g}{m_V/g} = \frac{5.6419}{1/\sqrt{\pi}} \approx 10$ (where $m_V / g = 1/\sqrt{\pi}$ is the vector meson mass in the $m \rightarrow 0$ limit) and volume $\vol=\mass \cdot\frac{N}{\sqrt{x}} \approx 100$, for several values of the lattice spacing $a = \frac{1}{g\sqrt{x}}$. For each value of $x$, the PDF is obtained according to \cref{eq:PDF_discrete} from $\mathcal{M}(\z)$ by a discrete Fourier transform. The results are shown in \cref{fig:PDF_x_dependence}.

\begin{figure}[tbp]
	\centering
	\def\figWidth{.89*\columnwidth}
	\def\figHeight{.3*\textwidth}
	\renewcommand{\tikzlinewidth}{0.2pt}
	\input{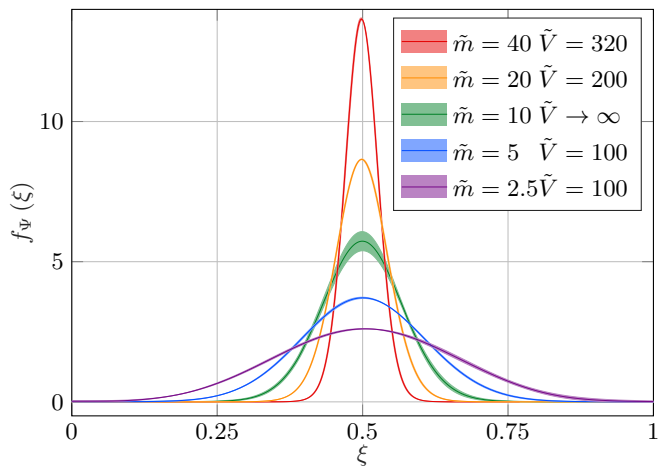}
	\renewcommand{\tikzlinewidth}{1.0pt}
	\caption{PDFs for different fermion masses $\mass = \frac{m \sqrt{\pi}}{g}$. The continuum limit $x \rightarrow \infty$ is taken with the volume $\vol = \mass \cdot \frac{N}{\sqrt{x}}$ kept fixed. The error bands denote the uncertainties due to the continuum extrapolations, and are smaller than the line width of the plot in most regions. For $\mass = 10$, instead, we estimate the errors due to finite $D$, $\Nt$, and $x$, and take the infinite volume limit. The error band in this case includes all uncertainties, dominated mostly by the volume effects. See main text and \appendixref{D} for details.}
	\label{fig:PDF_massdep}
\end{figure}

The periodic nature of the Fourier transform imposes a bound on the range of \bjorx that can be studied at finite $x$, given by $|\bjorx| \le \bjorx_\text{max} = \frac{\pi}{M} x$. Therefore, $x$ should be chosen large enough such that $\bjorx_\text{max} \ge 1$. Furthermore, the lattice spacing must be fine enough to resolve the oscillations of the matrix elements (see \cref{fig:matrixElement}). With these conditions met, the results are close to the continuum limit, allowing us to extrapolate the PDF to this limit (solid line in \cref{fig:PDF_x_dependence}) using simulation results at different values of $x$ (more details in \appendixref{D}). The volume must be large enough for the matrix element to vanish at large $|\z|$ within the system size. In this case, \bjorx in \cref{eq:PDF_discrete} can take real values $|\bjorx| \le \bjorx_\text{max}$, which results in a continuous curve for the PDF (as seen in the extrapolation in \cref{fig:PDF_x_dependence} and all lines in \cref{fig:PDF_massdep}).

The distribution function of an antifermion $\overline{\Psi}$~\cite{collins_2011} can be obtained from $\antipdf = - \pdfbare(-\bjorx)$. Due to the antisymmetry of the matrix elements and thus of \pdf, we numerically find $\pdf \equiv \antipdf$, which is expected in the continuum and thermodynamic limit due to charge conjugation symmetry (see \appendixref{C}). The average momentum fractions carried by fermions and antifermions are within $\braket{\bjorx} = \int_0^1 d\bjorx \left(\bjorx \pdfantipdf\right) = 0.50 \pm 0.02$ for all cases in \cref{fig:PDF_massdep}. Thus, fermions and antifermions each carry half of the hadron's momentum and both have the same PDF. These results agree with the expectations for a meson in the Schwinger model~\cite{SchwingerPartonPerturbative}.

We compare the PDF for the fermion mass $\mass = 10$ with the work of Mo and Perry~\cite{MoPerry}, where a finite basis in the two- and four-fermion approximation in light-front field theory was used. We find good agreement of the overall shape. Notably, our calculated $\pdf$, within error bands, remains real and non-negative for $0 \le \bjorx \le 1$, and vanishes for $|\bjorx| > 1$, all as required for distribution functions and the probability interpretation in the parton model. In contrast, the PDF calculated by Mo and Perry exhibits regions with negative values of $\pdf$ (see \cite{MoPerry} and inset on lower right in \cref{fig:PDF_x_dependence}).

Repeating the procedure for different values of the volume, we extrapolate the result for $\mass = 10$ to the thermodynamic limit (green curve in \cref{fig:PDF_massdep}). We carry out a systematic error analysis (more details in \appendixref{D}). The error band for the case $\mass = 10$ in the figure accounts for uncertainties due to $D$, $\Nt$, $x$ and $\vol$, with the latter two contributions dominating. We obtain a $6\%$ uncertainty at the peak $\bjorx = 0.5$.

Additionally, we compute the PDF for different fermion masses at fixed but large enough \vol. The results, already extrapolated to the continuum limit, are shown in \cref{fig:PDF_massdep}. We observe that larger volumes \vol are necessary as \mass increases in order to resolve more oscillations of the matrix elements that arise due to the steeper features of \pdf. The previous findings remain qualitatively unchanged across all masses, but the peak around $\bjorx = 0.5$ broadens for smaller \mass and sharpens for larger $\mass$. This is in good agreement with previous results~\cite{MoPerry,BERGKNOFF1977215,KROGER199858}, and with the limits of a step-function $\pdf \equiv 1$ for $0 \le \bjorx \le 1$ at $m=0$~\cite{SchwingerPartonPerturbative,SchwingerPartonNLO}, and a delta-function at $\bjorx=0.5$ when the fermion mass is infinite.

The PDFs for these additional masses are shown at finite volume, but extrapolated to the continuum limit, with error bands (mostly smaller than the line width) indicating the effect of uncertainties due to the extrapolation in $x$. Even though these PDFs are computed at fixed finite volume, since the value of $\vol$ is kept at least as large as in the simulations for the $\mass = 10$ case, we expect that finite-volume effects are similar. This can be tested with further data available in the future.

\phantomsection
\addcontentsline{toc}{section}{Conclusion}
{\sl Conclusion ---}
We have demonstrated how tensor networks can be used to directly compute PDFs from light-cone matrix elements in Minkowski space. In our approach, the light-front Wilson line becomes a stepwise evolution in space and time on the lattice that can be applied on the Hamiltonian eigenstate representing a hadron at rest. Tensor networks provide an efficient representation of eigenstates and enable simulating the necessary time evolution. This framework allows us to compute the PDFs of the fermion and the antifermion in the vector meson of the Schwinger model from first principles, yielding accurate results that fulfill the physical constraints. We present the PDFs for several fermion masses in the continuum limit. In previous TN calculations, the continuum limit of lattice gauge theories was only taken for time-independent properties. Our work extends this to a simulation including a time evolution for the first time, further establishing TNs for the direct calculation of dynamical quantities in lattice gauge theories.

Tensor network states can be applied for higher-dimensional lattice gauge theories. Even though the numerical costs are higher, several works have successfully demonstrated the applicability of projected entangled pair states (PEPS), multi-scale entanglement renormalization ansatz (MERA) and tree tensor networks (TTN) for lattice gauge theories, see \cite{banuls2020ropp,banuls2020epjd} for reviews. While the Schwinger model is a 1+1 dimensional model, our results demonstrate that light-cone physics can be directly addressed with TNs in a well-controlled way, and motivate further investigations in higher dimensions and with non-Abelian gauge theories towards QCD. Our study demonstrates that the relevant light-cone observables can be computed with moderate bond dimensions, which is a promising foundation for higher-dimensional systems. The stepwise evolution of the Wilson line can be adapted there as well, and static charges can be utilized to mimick the Wilson line if the Hilbert space is restricted to the physical space obeying the constraints of Gauss's law. Although further theoretical investigations are needed, notably to address renormalization effects~\cite{Korchemsky1987}, this would open the path on spin- and transverse-momentum-dependent partonic functions, as well as gluon PDFs.

As a related and complementary approach, tensor network formulations like the one presented in this work can also be realized on quantum simulation platforms~\cite{banuls2020epjd}. The hadronic state that was represented by a tensor network in our work can be transformed into a quantum circuit which prepares the initial state on a quantum computer, see for example~\cite{Guo2025} and references therein. The stepwise evolution of the Wilson line through a time evolution with a Hamiltonian containing static charges can be readily implemented with established time-evolution techniques on quantum devices~\cite{Davoudi2023,Farrell2024,Hariprakash2025}.

\phantomsection
\addcontentsline{toc}{section}{Acknowledgments}
{\sl Acknowledgments ---}
We thank Hsiang-nan Li and Enrique Rico for very helpful discussions. 
This work was partly supported by  the DFG (German Research Foundation) under Germany's Excellence Strategy -- EXC-2111 -- 390814868, and Research Unit FOR 5522 (grant nr. 499180199), and by the EU-QUANTERA project TNiSQ (BA 6059/1-1), as well as Taiwanese NSTC grants 113-2119-M-007-013, 112-2811-M-A49-543-MY2, 112-2112-M-A49 -021-MY3 and 113-2123-M-A49-001. K.C.\ is supported by the National Science Centre (Poland) grant OPUS No.\ 2021/43/B/ST2/00497.

\nocite{Banks1976,Crewther1980,Irving1983,Hamer1997,Schwinger_massSpectrum,Schwinger_QC,henneaux1992quantization,staggeredFermions,Schwinger_latticeMass,MassRenormalizationNumerical,MassRenormalizationNumerical_PoS,Zohar2015,Schwinger_massSpectrum,Schwinger_thermalEvolution}	
\bibliography{references} 

\clearpage
\phantomsection
\addcontentsline{toc}{section}{Supplemental Material}
\label{supplemental}

\section{Appendix A: Hamiltonian formulation of the Schwinger model\label{app:A}}
In the following, we introduce our conventions, and explain the main steps to obtain the spin model Hamiltonian~\cite{Banks1976,Crewther1980,Irving1983,Hamer1997,Schwinger_massSpectrum,Schwinger_QC} from the {massive Schwinger model Lagrangian including static charges. We start from the Lagrange density in \eqLagrangian but include the coupling to an external current $j^{\mu}$:
\begin{equation}
	\mathcal{L} = \overline{\Psi} (i \slashed{\partial}- g \slashed{A }-m )\Psi-\frac{1}{4}F_{\mu\nu}F^{\mu \nu} - A_{\mu} j^{\mu} .
\end{equation}
We consider a static charge density \Qdense with external current $j^{\mu} = (\Qdense, 0)$. The Hamiltonian density follows from a Legendre transformation (see~\footnote{We use Einstein summation convention, the slashed notation $\slashed{A} = \gamma^\mu A_\mu$ with $\gamma^0 = \sigma_z = \begin{pmatrix}1&0\\0&-1\end{pmatrix}$ and $\gamma^1 = i \sigma_y = \begin{pmatrix}0&1\\-1&0\end{pmatrix}$, $\overline{\Psi} = \Psi^\dagger \gamma^0$, and $\sigma^{\pm} =\frac{1}{2}\left(\sigma^{x}\pm i\sigma^{y}\right) = \frac{1}{2} \begin{pmatrix}    0	&	1 \pm 1\\  1 \mp 1	&	0  \end{pmatrix}$.} for our choice of $\gamma$-matrices:
\begin{align}
	{\mathcal{H}}_{\mathrm{Sch}} =& -i \bar{\Psi} \gamma^1 (\partial_1 + i g A_1) \Psi + m \bar{\Psi} \Psi + \frac{1}{2} E^2 \nonumber \\
	&+ (\partial_1 A_0) E	+ A_0 \left(g \conj{\Psi} \Psi + \Qdense \right) .
\end{align}
The electric field is defined as $E = F_{0,1} = \partial_0 A_1 - \partial_1 A_0$. The momentum that is conjugate to $A_0$ vanishes, $\frac{\partial \mathcal{L}}{\partial(\partial_0 A_0)} = 0$. Therefore, $A_0$ is not a dynamical degree of freedom but has the role of a Lagrange multiplier~\cite{henneaux1992quantization}, which enforces Gauss's law~\footnote{The full Lagrangian or Hamiltonian is the spatial integral over the corresponding density. We use $\int dz (\partial_1 A_0) E = \int dz [\partial_1 (A_0 E) - A_0 \partial_1  E]$ with the first term vanishing for $A_0 E = 0$ at the boundaries.}:
\begin{equation}
	\partial_1 E = g \conj{\Psi} \Psi + \Qdense .
\end{equation}
Consequently, without loss of generality, we can make use of the temporal gauge $A_0 = 0$ in the following, with Gauss's law as an additional constraint.

For a lattice formulation, we use Kogut-Susskind staggered fermions~\cite{staggeredFermions}, with fermionic fields
\begin{equation}
	\frac{1}{\sqrt{a}} \phi_{n} \equiv
	\left\{\begin{array}{lr}
	\Psi_\text{upper}(n a)~\text{if}~n~\text{even} \\
	\Psi_\text{lower}(n a)~\text{if}~n~\text{odd},
	\end{array}\right.
\end{equation}
containing only the upper or lower components of the Dirac spinors. The Hamiltonian becomes
\begin{align}
	H_{\mathrm{latt}}=&-\frac{i}{2 a}\sum_n \left (  \phi^{\dagger}_n e^{i\theta_n}  \phi_{n+1} - \phi^{\dagger}_{n+1} e^{-i\theta_n}  \phi_{n}\right ) \nonumber\\
	&+m_\text{lat}\sum_n(-1)^n\phi_n^{\dagger}\phi_n
	+\frac{a g^2}{2} \sum_n L_n^2.
	\label{eq:Hferm}
\end{align}
Here, $m_\text{lat} = m - \frac{ag^2}{8}$ contains the additive renormalization~\cite{Schwinger_latticeMass,MassRenormalizationNumerical,MassRenormalizationNumerical_PoS}. We introduce the variables $\theta_n \equiv ag A_{1}(n a)$ and $L_n \equiv \frac{1}{g} E(n a)$, and impose the canonical commutation relation 
\begin{equation}
	[\theta_n,L_m]=i\delta_{n,m}
	\label{eq:cancon}
\end{equation}
to quantize the system. In a ladder space, the operators can be represented as angular momentum operator $L \Ket{l} = l \Ket{l}$ and rising or lowering operators $e^{\pm i \theta} \Ket{l} = \Ket{l \pm 1}$~\cite{Banks1976,Irving1983}. Thus, $L_n$ measures the electric flux on a link from site $n$ to site $n+1$, while the gauge link $\W_n = e^{i \theta_n}$ lowers the electric flux on that link by one unit.

We apply a residual gauge transformation
\begin{equation}
	\Tilde{\phi}_n \equiv \phi_n \prod_{k \ge n}\left(e^{i \theta_{k}}\right) 
	\label{eq:residualtrafo},
\end{equation}
and a Jordan Wigner transformation
\begin{equation}
	\Tilde{\phi}_n \equiv \prod_{k<n}\left(-i\sigma_{k}^{z}\right) \sigma_n^{-}. \label{eq:JordanWigner}
\end{equation}
Furthermore, we subtract the ground-state energy $E_0 = - \sum_{n~\text{odd}} m$ of the strong coupling limit and scale the Hamiltonian by a factor $\frac{2}{ag^2}$ to obtain
\begin{align}
	H_{\mathrm{spin}} =& x \sum_n \left (  \sigma^{+}_n \sigma^{-}_{n+1} + \sigma^{-}_{n+1} \sigma^{+}_{n}\right ) \nonumber\\
	&+\frac{\mu}{2}\sum_{n} \left [ 1 + (-1)^n \sigma_n^z \right ]
	+ \sum_n L_n^2,
	\label{eq:Hspin_with_Ln}
\end{align}
with $\mu \equiv \frac{2}{ag^2} m_{\text{lat}}$ and $x \equiv \frac{1}{a^2g^2}$.

Physical states need to fulfill Gauss's law, which becomes on the lattice
\begin{equation}
	L_n-L_{n-1}=Q_n
\end{equation}
where $Q_n$ is the charge in the staggered prescription~\cite{Zohar2015,Hamer1997},
\begin{align}
	Q_n =& \phi_n^{\dagger}\phi_n-\frac{1}{2}\left[1-(-1)^n\right] + \Q_n \nonumber\\
	=& \frac{1}{2}\left[(-1)^n + \sigma^z_n\right]+ \Q_n ,
\end{align}
with $\Q_n$ being the external (static) charge on vertex $n$.

The Hilbert space can be restricted to the physical states only, expressing the non-dynamical gauge degrees of freedom in terms of the spin variables by summing all the charges to the left of a gauge link. After the Jordan-Wigner transformation, the electric flux operator in the physical Hilbert space reads
\begin{equation}
	L_n = \frac{1}{2}\sum_{k \le n} \left((-1)^k+\sigma_k^{z} + 2 \Q_k\right).
\end{equation}
Inserting this in \cref{eq:Hspin_with_Ln} leads to the Hamiltonian in \eqH.

\section{Appendix B: Wilson line in the physical subspace\label{app:B}}
Let us define $U(\Delta t)$ to be the evolution operator with the (fermionic) lattice Schwinger Hamiltonian in \cref{eq:Hferm} for a time interval $\Delta t$:
\begin{equation}
	U(\Delta t)=e^{-i \Delta t H}.
\end{equation}
The operator whose matrix element we want to compute, for the line going from $(m,t_m)$ to $(n,t_n)$ and assuming $n > m$, is of the form
\begin{equation}
	\phi^{\dagger}_n(t_n) e^{-i\theta_{n-1}(t_{n-1})} e^{-i\theta_{n-2}(t_{n-2})} \ldots e^{-i\theta_{m}(t_m)} \phi_m(t_m).
	\label{eq:pdfoperator}
\end{equation}
Each time argument indicates the Heisenberg picture evolved operator, for instance
\begin{equation}
	\phi_n(t_n)=U^{\dagger}(t_n) \phi_n U(t_n).
\end{equation}
Therefore, for the Wilson line (without loss of generality let us set $t_m = 0$), we can write
\begin{align}
	W(m\to n)
	=&
	U^{\dagger}(t_{n-1}) e^{-i\theta_{n-1}}
	U(\Delta t) e^{-i\theta_{n-2}} U(\Delta t) \nonumber\\
	&\ldots  e^{-i\theta_{m+1}} U(\Delta t) e^{-i\theta_{m}},
\end{align}
where $\Delta t$ corresponds to the evolution between the steps. With the definition
\begin{equation}
	\Tilde{\theta}_s = \sum_{k = m}^s \theta_{k},
\end{equation}
we can rewrite the Wilson line as
\begin{align}
	W(m\to n) 
	=&
	U^{\dagger}(t_{n-1}) \cneg{n-1} \cpos{n-2} \nonumber\\
	&U(\Delta t) \cneg{n-2} \cpos{n-3} U(\Delta t)) \nonumber\\
	&\ldots \cneg{m+1} \cpos{m} U(\Delta t)) \cneg{m} \\
	=&
	U^{\dagger}(t_{n-1}) \cneg{n-1} U_{n-2}(\Delta t) U_{n-3}(\Delta t) \nonumber\\
	&\ldots U_{m+1}(\Delta t) U_{m}(\Delta t)
	,\label{eq:wilsoncharges}
\end{align}
where 
\begin{equation}
	U_s(\Delta t)= \cpos{s} U(\Delta t) \cneg{s}
\end{equation}
is the evolution operator for one step $\Delta t$ transformed by the partial string from the starting point $m$ of the Wilson line to $s$. Using that $\theta$ and $L$ are canonically conjugate variables (see \cref{eq:cancon}),
the action of the string on the evolution operator reduces to transforming the electric flux terms in the Hamiltonian as
\begin{equation}
	e^{i \theta_n} L_n e^{-i \theta_n}=L_n-1.
	\label{eq:transfL}
\end{equation}
This means that $U_s(\Delta t)$ is the evolution operator with a modified Hamiltonian, where all links from $m$ to $s$ \emph{see} a displaced flux.

In our implementation, the fermionic operators contain flux strings going to the right boundary (see \cref{eq:residualtrafo}). These can be canceled by inserting pairs $e^{i \sum_{k \ge m} \theta_k} e^{-i \sum_{k \ge m} \theta_k}$ between pairs of $U_s$ and defining a new time evolution operator with a displaced flux
\begin{align}
	\Tilde{U}_s(\Delta t) =& e^{-i \sum_{k \ge m} \theta_k} \cpos{s} U(\Delta t) \cneg{s} e^{i \sum_{k' \ge m} \theta_k'} \nonumber\\
	=& e^{-i \sum_{k > s} \theta_k} U(\Delta t) e^{i \sum_{k' > s} \theta_k}.
\end{align}
Then, the operator in \cref{eq:pdfoperator} becomes
\begin{equation}
	U^{\dagger}(t_{n}) \Tilde{\phi}^{\dagger}_n \Tilde{U}_{n-1}(\Delta t) \ldots \Tilde{U}_{m+1}(\Delta t) \Tilde{U}_{m}(\Delta t) \Tilde{\phi}_m.
\end{equation}
An evolution with the modified Hamiltonian thus corresponds to stopping the flux line from the initial fermion operator with the static charge. The operators $e^{i \sum_{k \ge s} \theta_k}$ and $e^{-i \sum_{k \ge s} \theta_k}$ have the same effect as increasing or decreasing the static charge $\Q_s$ by one unit, respectively. This formalism allows us to calculate matrix elements without direct access to the gauge degrees of freedom of the state in the spin description, by moving a static charge stepwise in each zigzag step. Note that the evolution in \eqMatrixElements corresponds to two lattice sites per time evolution step, with static charges always at odd sites, and that we start with a time evolution followed by a spatial evolution.

\section{Appendix C: Symmetry of matrix elements\label{app:C}}
We consider the matrix elements of the form
\begin{equation}
	\mathcal{M}_{(t,x) \leftarrow (0,0)} = \braket{P| \bar{\psi}(t,x) W_{(t,x) \leftarrow (0,0)} \Gamma \psi(0,0)|P}
\end{equation}
as in \eqPDF, for any given matrix $\Gamma$ that satisfies $\Gamma^\dagger = \gamma^0 \Gamma \gamma^0$. From translational invariance in space and time, it follows:
\begin{align}
	\mathcal{M}&_{(t,x) \leftarrow (0,0)} \nonumber\\
		&= \braket{P| \bar{\psi}(0,0) W_{(0,0) \leftarrow (-t,-x)} \Gamma \psi(-t,-x)|P} \nonumber\\
		&= \left[ \braket{P| \psi^\dagger (-t,-x) W^\dagger_{(0,0) \leftarrow (-t,-x)} \Gamma^\dagger \gamma_0 \psi(0,0)|P} \right]^* \nonumber\\
		&= \left[ \braket{P| \bar{\psi} (-t,-x) W_{(-t,-x) \leftarrow (0,0)} \Gamma \psi(0,0)|P} \right]^* \nonumber\\
		&= \mathcal{M}^*_{(-t,-x) \leftarrow (0,0)}
\end{align}

Therefore, the real part of $\mathcal{M}$ is symmetric with respect to reflection across the origin $(0,0)$, while the imaginary part is antisymmetric.

Considering a hadron that obeys charge symmetry, $\pdf \equiv \antipdf$, one can utilize $\antipdf = - \pdfbare(-\bjorx)$ and the variable transformation $z \rightarrow -z$ to obtain
\begin{align}
    \antipdf =& - \int_{-\infty}^{\infty} \frac{dz^{-}}{4\pi} e^{i \bjorx P^+ z^-}\hspace{-1.61365pt} \mathcal{M}(z^-) \nonumber\\
    =& - \int_{-\infty}^{\infty} \frac{dz^{-}}{4\pi} e^{-i \bjorx P^+ z^-}\hspace{-1.61365pt} \mathcal{M}(-z^-) \nonumber\\
    \equiv \pdf =& \phantom{-} \int_{-\infty}^{\infty} \frac{dz^{-}}{4\pi} e^{-i \bjorx P^+ z^-}\hspace{-1.61365pt} \mathcal{M}(z^-).
\end{align}
This shows that $\mathcal{M}$ has to be antisymmetric with respect to the origin, and thus the symmetric real part has to vanish. Because we use open boundary conditions in our study, which break translational symmetry, small deviations from these expectations can be observed for finite lattice sizes. The symmetry properties and the vanishing real part of the matrix elements are recovered in the limit of infinite lattice size.

\section{Appendix D: systematic errors and extrapolations\label{app:D}}
Our approach to calculating PDFs in the Schwinger model involves discretizations and cutoffs. Here, we analyze the errors introduced by these approximations and discuss how they can be mitigated through extrapolations.
Generally, one could extrapolate in all parameters subsequently, as in the strategy employed in~\cite{Schwinger_massSpectrum}. Specifically, this requires extrapolating in $\Lcut$, $D$, \Nt, $x$, and $\vol$. If we were to consider only four data points for each extrapolation, this would necessitate $4^5 = 1024$ different parameter sets for each mass value. This would be feasible with sufficient computational time if high accuracies were required. Instead, we adopt a more resource-efficient method in this work.

Initially, we compute the PDF from matrix elements at a fixed point in the parameter space. We choose $L_\text{cut} = 10$, $D = 80$, $\Nt = 100$ and, for the mass $\mass = 10$ discussed here as an example, $x=100$ and $\vol=100$. We then systematically vary each parameter individually to assess the error that occurs when the corresponding parameter is kept finite. The final result is obtained by an extrapolation in one of the parameters, such as $\vol$ (for $\mass=10$ in \figpdfmassdep) or $x$ (for all other masses and in \figpdfxdep), which dominate the errors for our parameter choice (see \cref{fig:errors} for the contributions to the total error). We estimate the uncertainty for the case $\mass=10$ from the spread or extrapolations in the other parameters and combine them together with the uncertainty due to the extrapolation in \vol. The resulting error band can be seen in \figpdfmassdep. In all other cases ($\mass \ne 10$ and \figpdfxdep), we only take the continuum limit and the error bands correspond to the uncertainty of this extrapolation in $x$.

\begin{figure}[htbp]
	\centering
	\def\figWidth{.8*\textwidth/2}
	\def\figHeight{.3*\textwidth}
	\input{tikz/errors}
	\caption{Total error of the PDF for $\mass = 10$ in \figpdfmassdep and contributions from different error sources. The relative error at $\bjorx = \pm 0.5$ is $6\%$.}
	\label{fig:errors}
\end{figure}

\subsection{Bond dimension}
To assess the impact of the finite bond dimension on the matrix elements and PDFs, we perform simulations with $D \in \{40, 80, 120\}$. Accurate results can be obtained for $D \ge 40$. Our simulations show that the matrix elements vary on the order of $10^{-3}$ if scaled by the maximum amplitude of the matrix elements. For the PDF, each data point varies at most by $10^{-3}$ in absolute units. We set $D = 80$ in all subsequent simulations, making the error due to the finite bond dimension sub-leading (see \cref{fig:errors} for a comparison of error sources). We estimate the error $\epsilon_{D}$ as the maximum deviation between the results obtained with $D = 80$ and those with $D \in \{40, 120\}$. This estimation is done for each value of \z and \bjorx for the matrix elements and PDFs, respectively. We note that $D$ also serves as the truncation parameter for the MPS after each time evolution step, with $\chi = D$ in our simulations.

\subsection{Trotterization}
In each zigzag step in the evolution of the Wilson line, the state is evolved by a time step $\dt$, further divided into $\Nt$ Trotter steps of size $\dtrotter$. The exact time evolution is recovered in the $\Nt \rightarrow \infty$ limit. To examine the dependence on \Nt, we conduct simulations with a range of values for \Nt, as shown in \cref{fig:tau_dependence}. The real part of the matrix elements vanishes as $\Nt \rightarrow \infty$ (\cref{sfig:M_real_tau_dependence}), while the imaginary part converges to an oscillating function that vanishes for large distances (\cref{sfig:M_imag_tau_dependence}).

We extrapolate to $\Nt \rightarrow \infty$, but note that these extrapolations are only used to estimate the error due to a finite $\Nt$ in the central value of the PDF quoted in this work. We apply three different fit functions (all extrapolation parameters $c_i$ are real):
\begin{enumerate}
	\item[\textbf{fit1}] Because of the second-order Suzuki-Trotter decomposition employed in our simulation, the error is expected to be of order $\Nt^{-2}$. We therefore apply a second-order fit to the real part of the matrix elements for each value of \z independently, $\Re(\cor(\z)) = c_1(\z) + c_2(\z) \cdot \Nt^{-2}$. The imaginary part is sensitive to higher orders, and we fit $\Im(\cor(\z)) = c_3(\z) + c_4(\z) \cdot \Nt^{-2} + c_5(\z) \cdot \Nt^{-4}$.
	\item[\textbf{fit2}] The trotterized Hamiltonian remains unitary, and we find that each time evolution step by $\dtrotter$ introduces an error in the form of a phase factor. We therefore fit the function $\cor(\z) = c_6(\z) e^{i c_7 \Nt^{-2} \z}$. All complex values $\cor(\z)$ are utilized simultaneously to fit the $\frac{N}{2}$ values $c_6(\z)$ and one global phase factor $c_7$.
	\item[\textbf{fit3}] Similarly to fit2, but allowing $c_7$ to vary with \z.
	The resulting phase factors for different \z agree with the value obtained from fit2 (see e.g. legend entries of subfigures c) - f) in \cref{fig:tau_dependence})~\footnote{We only find deviations from this for large $|\z|$, where the matrix elements become small and numerical errors dominate.}. This confirms the validity of the ansatz in fit2.
\end{enumerate}

For subsequent computations, we set $\Nt = 100$ and estimate the error arising from this finite choice as $\epsilon_{\Nt} = \max |\cor(\Nt > 100) - \cor(100)|$. This corresponds to the maximum distance between the matrix element at $\Nt = 100$ and those with larger $\Nt$, including the extrapolated values with $\Nt \rightarrow \infty$. Symmetric errors are assumed: $\cor = \cor(\Nt = 100) \pm \epsilon_{\dtrotter}$. With this procedure, we can estimate the error in a conservative way for the data with finite \Nt = 100. The results are shown as error bars in \cref{fig:tau_dependence} on the matrix element at $\Nt = 100$.

\subsection{Truncation of electric flux}
The time evolution operator for the electric part of the Hamiltonian is implemented as a matrix product operator (MPO)~\cite{Schwinger_thermalEvolution}, which is restricted to changes in the electric flux in the range $-\Lcut$ to $\Lcut$.
Varying $\Lcut$ allows us to extrapolate to $\Lcut \rightarrow \infty$ and to estimate the error $\epsilon_{\Lcut}$ due to the finite truncation $\Lcut = 10$ that we use in all other calculations, similarly to the previous analysis of the error caused by \Nt. The estimate is the maximum difference between the value with $\Lcut = 10$ and the values with $\Lcut > 10$, including extrapolations $\Lcut \rightarrow \infty$ with three different fit models (linear or quadratic in $\frac{1}{\Lcut}$, or both). \Cref{fig:Lcut_dependence} shows an example of the dependence of a matrix element on \Lcut. We generally find that a cutoff $\Lcut = 1$ leads to qualitatively incorrect results, while $\Lcut = 2$ reproduces the PDF with errors of the order $10^{-4}$. The errors for $\Lcut = 10$ are further reduced by one order of magnitude, making the effect of the cutoff completely negligible in the total error budget (see also \cref{fig:errors} for a comparison of error sources).

\subsection{Continuum limit}
The continuum limit corresponds to $a = \frac{1}{g\sqrt{x}} \rightarrow 0$, or equivalently $x \rightarrow \infty$. To study this limit, we keep the volume $\vol = \mass \cdot \frac{N}{\sqrt{x}}$ fixed and vary $x$. 
Since matrix elements at different $x$ are not evaluated at the same physical distances $\frac{\z}{2\sqrt{x}}$, we extrapolate the \pdf rather than the matrix elements. The PDF and examples of the extrapolations are shown in \cref{fig:x_dependence}. We apply linear and quadratic fits in the lattice spacing. Similarly to the analysis of the Trotter error, the maximal difference of the PDF from the point $x =100$ for all extrapolated values and all points with $x > 100$ is taken as an estimate of the error $\epsilon_{x}$ due to a finite $x$ in our simulations (green error bars at $\frac{1}{\sqrt{x}} = 0.1$ in \cref{fig:x_dependence}). It contributes to the total error (see \cref{fig:errors}) and, therefore, to the error band for $\mass = 10$ in \figpdfmassdep.

Instead of estimating the errors, we can also take the continuum limit at fixed \vol. We take the maximum $f_\text{max}$ and the minimum $f_\text{min}$ among the extrapolations and the value at $x = 100$. The continuum limit is taken to be the central value $\frac{f_\text{max} + f_\text{min}}{2}$ with errors $\pm \frac{f_\text{max} - f_\text{min}}{2}$ (black pentagons with error bars at $\frac{1}{\sqrt{x}} = 0$ in \cref{fig:x_dependence}). This procedure is applied to obtain the continuum extrapolations in \figpdfxdep for $\mass = 10$ at a fixed volume $\vol =  100$, and in \figpdfmassdep for all masses except $\mass = 10$ where an extrapolation to $\tilde{V} \rightarrow \infty$ is also performed.

\subsection{Infinite volume limit}
We study the volume dependence in a way similar to the continuum limit. The number of lattice sites $N$ is varied while the other parameters are kept fixed. We extrapolate to $N \rightarrow \infty$ using linear and quadratic fits in $\frac{1}{N}$, as shown in \cref{fig:N_dependence}. We take the minimum and maximum among the fits and the result with $N = 100$ which was used for the estimate of the other error sources. The PDF in the infinite volume limit is the midpoint between these extremal values (see black pentagons in \cref{fig:N_dependence} and central value of the PDF with $\mass = 10$ in \figpdfmassdep). The error $\epsilon_{\vol}$ corresponds to the distances to the minimum or maximum (see error bars in \cref{fig:N_dependence}). The same strategy shows that the imaginary part of \pdf vanishes within the errors. We currently only extrapolate to the infinite volume limit for $\mass = 10$. For this case, the central value of the extrapolation is shown in \figpdfmassdep, with error bars corresponding to the combined error of the infinite-volume extrapolation and all other error estimates.

\subsection{Total error}
The estimated errors due to finite $\Lcut$, $D$, $\Nt$, $x$ and $\vol$ are combined by quadrature to obtain an uncertainty band for the PDF in \figpdfmassdep for $\mass = 10$ (for all other masses, the error band corresponds only to the uncertainty of the continuum limit extrapolation). The final result is $\pdfbare = \pdfbare(\vol \rightarrow \infty) \pm \epsilon$ with $\epsilon = \sqrt{|\epsilon_{D}|^2 + |\epsilon_{\Nt}|^2 + |\epsilon_{\Lcut}|^2 + |\epsilon_{x}|^2 + |\epsilon_{\vol}|^2}$. A comparison of error sources in \cref{fig:errors} shows that finite-volume and lattice-discretization errors dominate. The Trotter step size and bond dimension truncation have only subleading effects, while that of the electric field cutoff \Lcut is negligible.
\clearpage
\begin{figure*}[htbp]
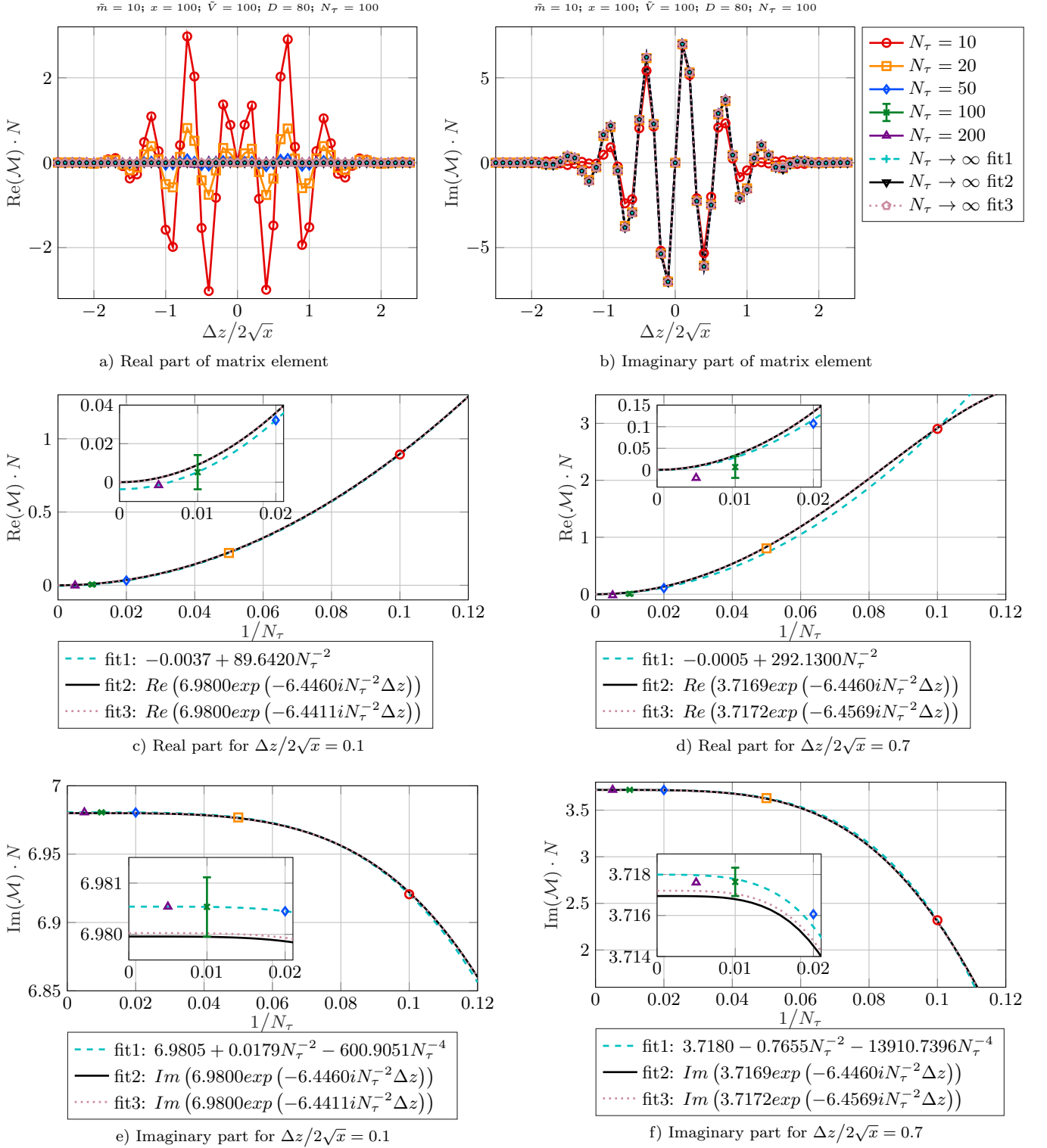

	\centering
	\def\figWidth{.35*\textwidth}
	\def\figHeight{.265*\textwidth}
	\subfloat[Real part of matrix element\label{sfig:M_real_tau_dependence}]{%
		\input{tikz/M_real_tau_dependence}
	}\hfill
	\subfloat[Imaginary part of matrix element\label{sfig:M_imag_tau_dependence}]{%
		\input{tikz/M_imag_tau_dependence}
	}\hfill\\
	\def\figWidth{\textwidth/2.5}
	\def\figHeight{\textwidth/5}
	\subfloat[Real part for $\z \big/ 2\sqrt{x} = 0.1$\label{sfig:M_real_tau_extrapolation_d2}]{%
		\input{tikz/M_real_tau_extrapolation_d2}
	}\hfill
	\subfloat[Real part for $\z \big/ 2\sqrt{x} = 0.7$\label{sfig:M_real_tau_extrapolation_d14}]{%
		\input{tikz/M_real_tau_extrapolation_d14}
	}\hfill
	\subfloat[Imaginary part for $\z \big/ 2\sqrt{x} = 0.1$\label{sfig:M_imag_tau_extrapolation_d2}]{%
		\input{tikz/M_imag_tau_extrapolation_d2}
	}\hfill
	\subfloat[Imaginary part for $\z \big/ 2\sqrt{x} = 0.7$\label{sfig:M_imag_tau_extrapolation_d14}]{%
		\input{tikz/M_imag_tau_extrapolation_d14}
	}
	\caption{Dependence of the matrix elements on the number of Trotter steps $\Nt$. Three different fit models are applied to extrapolate to $\Nt \rightarrow \infty$; fit1 applies a polynomial fit for the real and imaginary parts of the matrix elements independently; fit2 assumes an exponential form with two parameters. A global fit is applied for all distances \z; fit3 makes use of the same exponential form, but the fit is done for each value of \z independently. The error bar at $\Nt = 100$ is estimated as the maximal deviation from the extrapolations. See main text for further details.}
	\label{fig:tau_dependence}
\end{figure*}

\begin{figure*}[htbp]
	\centering
	\def\figWidth{.5*\textwidth}
	\def\figHeight{.265*\textwidth}
	\input{tikz/M_imag_Lcut_extrapolation_d14}
	\caption{Dependence of the Matrix elements on the truncation of the electric flux \Lcut. Shown is the imaginary part for $\z \big/ 2\sqrt{x} = 0.7$. Only values with $\Lcut \ge 4$ are used in the extrapolations. The error bar at $\Lcut = 10$ is estimated as the maximal deviation from the extrapolations.}
	\label{fig:Lcut_dependence}
\end{figure*}

\begin{figure*}[htbp]
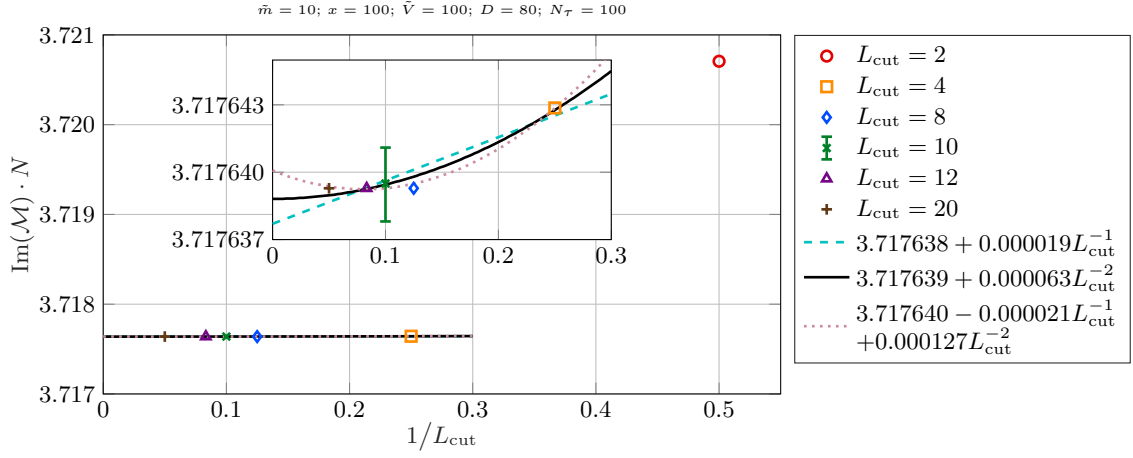
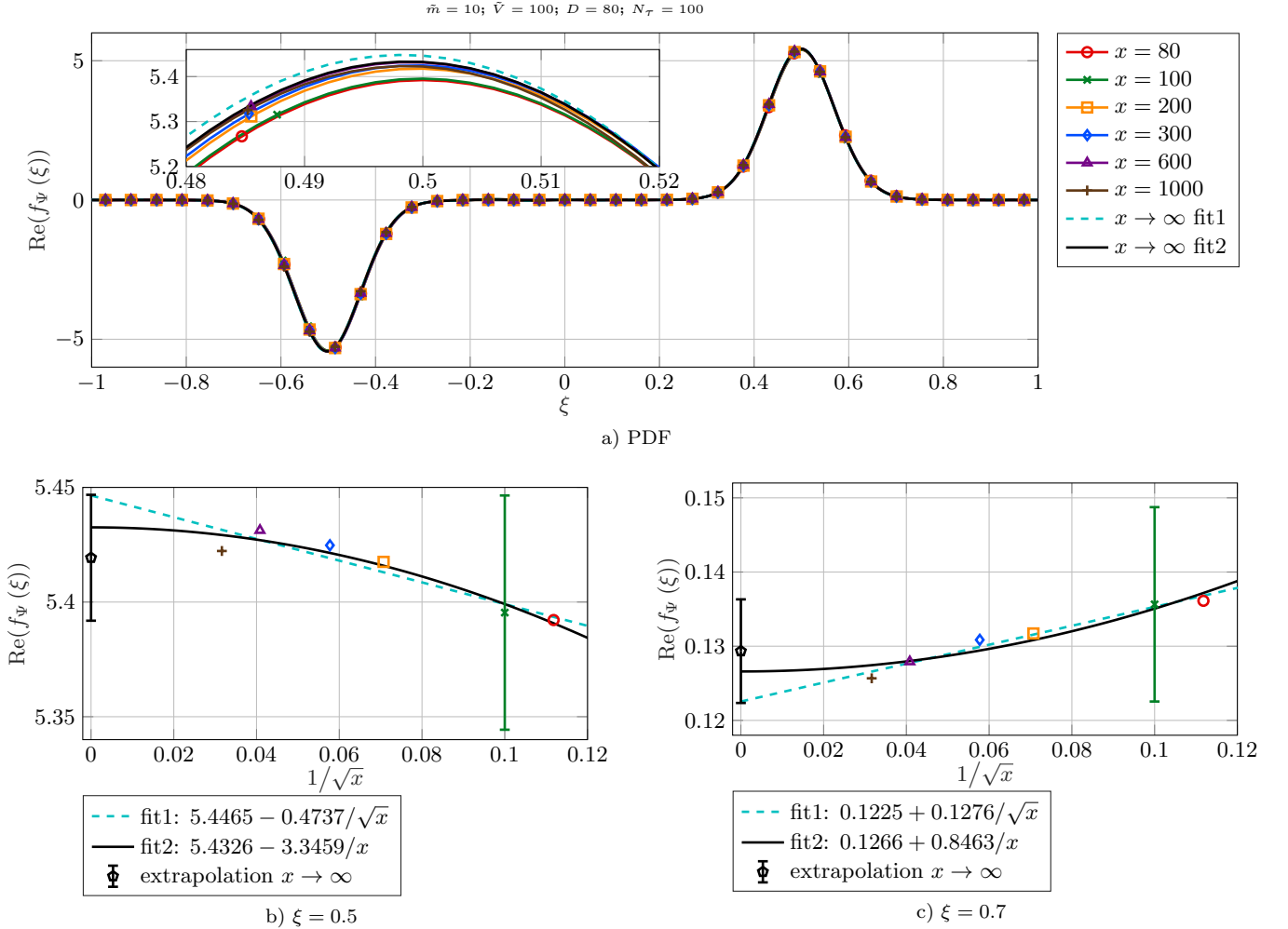

	\centering
	\def\figWidth{.75*\textwidth}
	\def\figHeight{.265*\textwidth}
	\subfloat[PDF\label{sfig:PDF_real_x_dependence}]{%
		\input{tikz/PDF_real_x_dependence}
	}\hfill
	\def\figWidth{\textwidth/2.5}
	\def\figHeight{\textwidth/5}
	\subfloat[$\bjorx = 0.5$\label{sfig:PDF_real_x_extrapolation_bjorx5}]{%
		\input{tikz/PDF_real_x_extrapolation_bjorx5}
	}\hfill
	\subfloat[$\bjorx = 0.7$\label{sfig:PDF_real_x_extrapolation_bjorx7}]{%
		\input{tikz/PDF_real_x_extrapolation_bjorx7}
	}\hfill
	\caption{Dependence of the PDF on $x$. Two fit models are applied for extrapolating to $x \rightarrow \infty$; fit1: linear in lattice spacing $a$, fit2: quadratic in $a$. The green error bar at $\frac{1}{\sqrt{x}} = 0.1$ is estimated as the maximal deviation from the extrapolations, and quantifies the uncertainty for simulations at a finite $x = 100$. The black pentagon with error bars corresponds to our continuum extrapolation. See main text for details.}
	\label{fig:x_dependence}
\end{figure*}

\begin{figure*}[htbp]
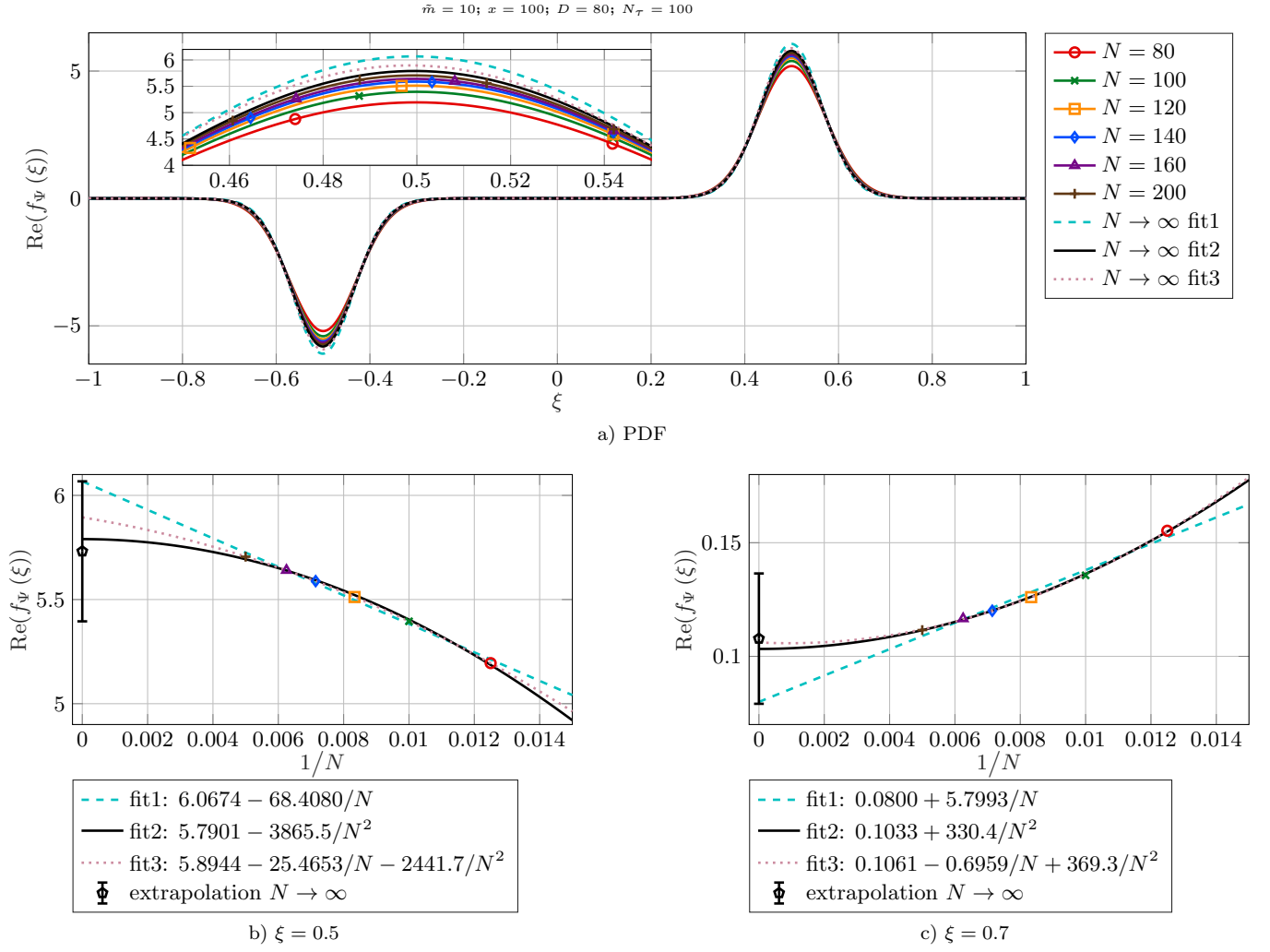

	\centering
	\def\figWidth{.75*\textwidth}
	\def\figHeight{.265*\textwidth}
	\subfloat[PDF]{%
		\input{tikz/PDF_real_N_dependence}
	}\hfill
	\def\figWidth{\textwidth/2.5}
	\def\figHeight{\textwidth/5}
	\subfloat[$\bjorx = 0.5$\label{sfig:PDF_real_N_extrapolation_bjorx5}]{%
		\input{tikz/PDF_real_N_extrapolation_bjorx5}
	}\hfill
	\subfloat[$\bjorx = 0.7$\label{sfig:PDF_real_N_extrapolation_bjorx7}]{%
		\input{tikz/PDF_real_N_extrapolation_bjorx7}
	}\hfill
	\caption{Dependence of the PDF on the volume $N$. Three different fit models are applied to extrapolate to $N \rightarrow \infty$; fit1: linear in $1/N$, fit2: quadratic in $1/N$, fit3: linear and quadratic terms. The black pentagon with error bars corresponds to our continuum extrapolation. See main text for details.}
	\label{fig:N_dependence}
\end{figure*}

\end{document}